\documentclass{aa}  

\usepackage{graphicx}
\usepackage{txfonts}
\usepackage[pdftex,dvipsnames]{xcolor}
\usepackage{rotating}
\usepackage{pdflscape}
\usepackage{hyperref}
\usepackage{afterpage}
\usepackage{float}

\begin{document}

   \title{Magnetar counterparts, kinematics and birth sites with HST and JWST}


   \author{A. A. Chrimes
          \inst{1}\fnmsep\inst{2}\thanks{ESA Research Fellows}, 
          J. D. Lyman\inst{3},\ 
          A. J. Levan\inst{2}\fnmsep\inst{3},
          A. Borghese\inst{4}$^{\star}$,\
          J. H. J. de Bruijne\inst{1},\
          A. S. Fruchter\inst{5},\
          M. G. Guarcello\inst{6},\ 
          C. Kouveliotou\inst{7,8},\
          N. R. Tanvir\inst{9}\
          and K. Wiersema\inst{10}\ 
          }

   \institute{European Space Agency (ESA), European Space Research and Technology Centre (ESTEC), Keplerlaan 1, 2201 AZ Noordwijk, the Netherlands \\
              \email{ashley.chrimes@esa.int}
         \and
            Department of Astrophysics/IMAPP, Radboud University, PO Box 9010, 6500 GL Nijmegen, The Netherlands
         \and
            Department of Physics, University of Warwick, Gibbet Hill Road, CV4 7AL Coventry, United Kingdom  
         \and
            European Space Agency (ESA), European Space Astronomy Centre (ESAC), Camino Bajo del Castillo s/n, 28692 Villanueva de la Cañada, Madrid, Spain 
         \and
            Space Telescope Science Institute, 3700 San Martin Drive, Baltimore, MD 21218, USA
         \and
            Istituto Nazionale di Astrofisica (INAF) – Osservatorio Astronomico di Palermo, Piazza del Parlamento 1, 90134 Palermo, Italy
         \and
            Department of Physics, The George Washington University, Corcoran Hall, 725 21st St NW, Washington, DC 20052, USA
         \and
            GWU/Astronomy, Physics and Statistics Institute of Sciences (APSIS)
         \and
            School of Physics and Astronomy, University of Leicester, University Road, Leicester, LE1 7RH, United Kingdom
         \and
            Centre for Astrophysics Research, University of Hertfordshire, College Lane, Hatfield, AL10 9AB, United Kingdom
             }

   \date{Received 12 March, 2026; accepted 20 June, 2026}

 
  \abstract
   {}
   {Magnetars are highly magnetised, isolated neutron stars with uncertain formation channels. They comprise a potentially significant fraction of the young neutron star population in the Milky Way, and are implicated in the explosion mechanisms of some of the most powerful explosions in nature. We aim to identify magnetars in the near-infrared with {\em Hubble} Space Telescope (HST) and {\em James Webb} Space Telescope (JWST) imaging, in order to measure their proper motions and search for their birth sites.}
   {Candidate infrared counterparts are selected based on variability, colours and proper motions which are outliers with respect to other sources in the field. Precise proper motions are obtained by tying HST/WCF3 and JWST/NIRcam images to the {\em Gaia} absolute astrometric reference frame.}
   {We newly identify infrared counterpart candidates for PSRJ1622$-$4950, 1RXSJ\,170849.0-400910 and CXOUJ164710.2-455216, representing a substantial increase in the population.
   The past trajectory of the 1RXSJ\,170849.0-400910-associated source coincides with the supernova remnant G346.6-0.2. The transverse velocity distribution of magnetars is found to be marginally inconsistent with that of young pulsars, due primarily to a dearth of high velocity magnetars. A candidate birth site is identified inside the cone of possible past trajectories in nearly every case. We show, based on the inferred kinematic ages, that magnetar characteristic ages may frequently be lower than the true age, but caution that this result depends on the reliability of the birth site associations.}
   {We conclude that magnetars are broadly similar in terms of their kinematics and birth sites to the wider Galactic neutron star population, consistent with magnetar formation being a common outcome of massive star core-collapse. However, tentative evidence for a dearth of high-velocity magnetars is emerging. If real, this may arise from physical differences in the progenitor population giving rise to magnetars, or from differences in their post-formation velocity evolution.}


   \titlerunning{Magnetar counterparts and kinematics with HST and JWST}
   \authorrunning{A. A. Chrimes et al.}
   
   \maketitle
%

\section{Introduction}

    Magnetars are isolated neutron stars with very high magnetic fields, typically 10--100 times stronger than pulsars. Due to their strong fields they spin down rapidly, and the population within the Milky Way exhibits slow rotation periods $P>1$s. Their electromagnetic emission exceeds that expected solely from their spin-down via magnetic braking, given $P$ and $\dot{P}$, and is thought to be additionally powered by the decay of their intense magnetic fields \citep[see e.g.][for reviews]{2015SSRv..191..315M,2015RPPh...78k6901T,2017ARA&A..55..261K,2021ASSL..461...97E,2021Univ....7..351I}. They have young characteristic ages of $<10^{5}$\,yr, as inferred from $P$ and $\dot{P}$ under the assumption of magnetic braking, a constant dipolar field and braking index $n=3$ \citep[e.g.][]{2015PhRvD..91f3007H}. Around $\sim$30 have been directly observed, predominantly in the plane of the Milky Way, plus a few in the Magellanic Clouds \citep{2014ApJS..212....6O}. 
    
    Most known magnetars have thus far been discovered through high-energy outbursts \citep{2018MNRAS.474..961C} or flares \citep[e.g.][]{1979Natur.282..587M,1999ApJ...510L.115K,2005Natur.434.1107P,2024FrASS..1188953N}, some of which can be luminous enough to be detected over extragalactic distances \citep{2021ApJ...907L..28B,2024Natur.629...58M}. Historically, magnetars were labelled as anomalous X-ray pulsars (AXPs) or soft gamma repeaters (SGRs) based on their observational properties, but are now understood to be manifestations of the same physical objects \citep[e.g.][]{1995A&A...299L..41V,1996ApJ...473..322T,1998Natur.393..235K}.

    Follow-up of Galactic magnetars often reveals multi-wavelength counterparts, from ubiquitous quiescent X-ray emission to several optical-infrared counterparts \citep[e.g.][]{2004ApJ...617L..53T,2005AdSpR..35.1177M,2008A&A...482..607T,2011MNRAS.416L..16D,2018ApJ...854..161L} and a handful of radio detections \citep{2006Natur.442..892C,2008ApJ...679..681C,2010ApJ...721L..33L,2013MNRAS.435L..29S,2020ApJ...896L..30E}.

    Magnetars are known to be variable at X-ray and optical/near-infrared (NIR) wavelengths. In some cases variations in the NIR flux appear to correlate with X-ray activity \citep[e.g.][]{2002Natur.417..527K,2004ApJ...617L..53T,2005A&A...438L...1I,2005MNRAS.363..609D,2025AandA...696A.127C}, in others not \citep[e.g.][]{2005ApJ...627..376D,2006ApJ...652..576D,2022ApJ...926..121L}. This variability is in any case a useful tool for identifying NIR counterparts: given crowded sight-lines in the Galactic plane and relatively large positional uncertainties from X-ray observations, identifying if one of a number of NIR candidates is variable is an effective way to unambiguously identify the counterpart \citep[e.g.][]{2002ApJ...580L.143I,2005A&A...438L...1I}.

    Magnetars are also typically redder than field stars at NIR wavelengths \citep[e.g.][]{2022MNRAS.512.6093C}. The optical-IR flux is likely non-thermal and magnetospheric in origin \citep[e.g.][]{2007Ap&SS.308..203M}, also given the correlation with X-ray emission seen in some cases, and the power-law spectral energy distribution (SED) seen in 4U0142$+$61 \citep{2016MNRAS.458L.114M,2024ApJ...972..176H}. However, this object appears spectrally variable on a timescale of years, and was previously observed to have a NIR bump in the SED consistent with blackbody emission from a debris disk \citep[][]{2006Natur.440..772W,2007ApJ...657..441E}. 
    
    Magnetars are often kinematically distinct with respect to the objects in the surrounding field \citep[e.g.][]{2007ApJ...662.1198H,2012ApJ...761...76T,2013ApJ...772...31T}. As neutron stars, they have likely received large natal kicks imparting velocities which are much greater than the velocity dispersion and bulk motion of field stars \citep[e.g.][]{2005MNRAS.360..974H,2017A&A...608A..57V,2020MNRAS.494.3663I,2024A&A...687A.272D,2025ApJ...989L...8D}. Determining the proper motions of magnetars, and hence their transverse velocities if a distance can be measured or estimated, allows us to determine their natal kick distribution, and ultimately their birth sites by tracing back their motion. This is possible thanks to their young ages. There are several examples of magnetars associated with supernova remnants (SNRs) and stellar clusters \citep[e.g.][]{2006ApJ...636L..41M,2007ApJ...667.1111G,2012ApJ...751...53A,2022ApJ...926..121L}; the latter also enables progenitor mass estimates based on the age of the cluster \citep{2009ApJ...707..844D,2010A&A...520A..48R}. 

    There are two main ideas for how magnetars obtained their extreme magnetic fields: either through magnetic flux conservation during the core-collapse process \citep[the `fossil field' hypothesis,][]{1964ApJ...140.1309W,2006MNRAS.367.1323F,2020MNRAS.495.2796S}, or through a dynamo upon neutron star formation \citep{1993ApJ...408..194T,2014MNRAS.445.3169O,2015Natur.528..376M,2022ApJ...926..111W,2022A&A...667A..94R}. The recent discovery of a Wolf-Rayet star with a magnetic field strong enough for magnetar formation suggests that the inherited fossil field scenario is plausible \citep{2023Sci...381..761S}, while the explanation of some high-energy extragalactic transients with (millisecond) magnetar central engines \citep[e.g. gamma-ray bursts and super-luminous supernovae,][]{2014MNRAS.438..240G,2015MNRAS.454.3311M,2024MNRAS.527.6455O, 2025ApJ...989L..41M} is in favour of their rapid rotation at birth. However, there is no evidence of additional energy injection into the SNRs surrounding Galactic magnetars due to post-birth spin down of a rapidly rotating neutron star \citep[][]{2006MNRAS.370L..14V}. Indeed, population synthesis studies suggest that the population of Galactic magnetars were likely not born with sufficient rotational energy to power such high energy transients \citep{2015ApJ...813...92R}. It is notable that not every Galactic magnetar has a clear SNR or cluster association \citep[e.g.][]{2013ApJ...772...31T,2025AandA...696A.127C}, despite their young characteristic (spin-down) ages.
    
    Another class of transients linked with magnetars are fast radio burst (FRBs), particularly since the detection of FRBs from the Galactic magnetar SGR1935+2154 \citep{2020Natur.587...54C,2020Natur.587...59B}. If the FRB-magnetar association is universal, the subsequent discovery of FRBs in a globular cluster \citep{2022Natur.602..585K} and ancient elliptical galaxies \citep{2023ApJ...950..175S} may point towards some fraction of magnetar formation through non-core-collapse channels with long delay times, such as the accretion or merger induced collapse of white dwarfs \citep{1991ApJ...367L..19N,2006MNRAS.368L...1L,2019ApJ...886..110M}. 

    Magnetar velocities and birth sites can be compared with the wider (young) neutron star population. In this way we can discern any differences in their progenitor systems, or otherwise \citep[e.g.][]{2025arXiv251106554H}. High resolution NIR imaging is ideally suited for this task. Measuring proper motions with even the highest resolution X-ray imaging is challenging \citep[but possible for high proper motion objects or over long baselines,][]{2009AJ....137..354K,2024ApJ...976..228R}, and only a handful of Galactic magnetars are radio-bright \citep[and hence amenable to radio interferometry,][]{2007ApJ...662.1198H,2012ApJ...748L...1D,2020MNRAS.498.3736D,2024ApJ...971L..13D}. Efforts have successfully been made with ground-based NIR observations assisted by adaptive optics \citep{2012ApJ...761...76T,2013ApJ...772...31T}, but challenging absolute astrometry and the sensitivity of such observations is a limitation. 
    
    In this paper, we identify near-infrared (NIR) magnetar counterparts in repeat {\em Hubble} Space Telescope (HST) imaging by searching for red, variable and kinematically distinct sources consistent with magnetar X-ray localisations. In one case, these observations are supplemented by {\em James Webb} Space Telescope (JWST) imaging. By measuring the proper motions of magnetar NIR counterparts with respect to previous imaging, new constraints on the velocity distribution, birth sites and kinematic ages of several Galactic magnetars are obtained. 

    The paper is structured as follows. In Section \ref{sec:obs} we describe the HST and JWST data used, the data reduction and the magnetars in our sample. Section \ref{sec:method} outlines in the methodology including photometry, astrometry, proper motion measurement and counterpart identification. Section \ref{sec:results} presents the results including three new magnetar counterparts, transverse velocities and birth site identifications. A discussion and conclusions follow in sections \ref{sec:discuss} and \ref{sec:conc}. Magnitudes are reported throughout in the AB system \citep{1983ApJ...266..713O}, and uncertainties are stated at 1$\sigma$ confidence unless otherwise stated.


\section{Observations and data reduction}\label{sec:obs}
\subsection{Hubble Space Telescope}
The new HST observations used in this work (program GO 17927, PI: Chrimes) are summarised in Table \ref{tab:obsv}. All observations were taken with the Wide Field Camera 3 \citep[WFC3][]{2008SPIE.7010E..1EK} and each observation consists of either 3 or 4 dithered exposures. The sample comprises 15 magnetars for which at least one candidate counterpart was identified in previous HST observations \citep{2022MNRAS.512.6093C}. Also listed are any previous epochs of HST observations, the filters used and the program IDs of those observations. We reduced the data following standard {\sc astrodrizzle} procedures \citep{2002PASP..114..144F}\footnote{Available with {\sc drizzlepac}, \url{https://www.stsci.edu/scientific-community/software/drizzlepac}}, including geometric distortion corrections. The calibrated {\sc \_flt} files for each image, which have undergone flat fielding and dark subtraction, were drizzled with a final orientation north up/east left, a {\sc pixfrac} of 0.8 and a final pixel scale of 0.065\,arsec\,pixel$^{-1}$.

\begin{table*}
	\centering
	\caption{Details of the {\em Hubble} Space Telescope observations taken under program 17927. Previous epochs of HST and/or JWST observations are listed in the final three columns. All observations use HST WFC3/IR and the F125W, F140W or F160W NIR filters, with the exception of the JWST/NIRcam observations of CXOUJ1647. For each magnetar there is one row per filter (including observations in GO 17927 and previous programs where available), unless there are multiple epochs of imaging in that filter. Blank entries indicate a lack of observations in that filter, either in the 2025-2026 epoch newly reported in this paper, or previous epochs. Shortened magnetar names are indicated in the first column, which are subsequently used throughout for brevity. The previous epoch data has been published by \citet[][program IDs 14805, 15348, 16019]{2022MNRAS.512.6093C}, \citet[][program IDs 14055, 14502, 16505]{2022ApJ...926..121L}, \citet[][program IDs 12306, 12672, 16019]{2025AandA...696A.127C}, \citet[][program IDs 14055, 14502]{2018ApJ...854..161L} and \citet[][program ID 1905]{2025A&A...693A.120G}. }
	\label{tab:obsv}
	\begin{tabular}{lllclcll} 
		\hline
		  \hline
            Magnetar	&	Epoch	&	Filter & Start date/time	&	Exptime	& Previous	&	Filter of & Program ID of 	\\
                 	&		      &            & [UT]	&	 [s]	& 	epoch(s)	&	prev. epoch(s)	 & prev. epoch(s)	\\
            \hline													
            PSRJ1622(-4950)	&	2025.2	&	F125W	&	2025-02-28 06:26:40	&	1058.819	&	2020.7	&	F125W	&	16019	\\
            	&	2025.2	&	F160W	&	2025-02-28 06:47:25	&	1058.819	&	2020.7	&	F160W	&	16019	\\
            SWIFTJ1822(.3-1606)	&	2025.2	&	F125W	&	2025-03-30 19:53:25	&	1058.819	&	2018.5	&	F125W	&	15348	\\
            	&	2025.2	&	F160W	&	2025-03-30 20:14:10	&	1058.819	&	2018.5	&	F160W	&	15348	\\
            SWIFTJ1834(.9-0846)	&	2025.2	&	F125W	&	2025-03-30 21:27:23	&	1058.819	&	2020.2	&	F125W	&	16019	\\
            	&	2025.2	&	F160W	&	2025-03-30 21:48:08	&	1058.819	&	2020.2	&	F160W	&	16019	\\
            SGR1627(-41)	&	2025.3	&	F125W	&	2025-04-24 14:32:17	&	1058.819	&	2020.7	&	F125W	&	16019	\\
            	&	2025.3	&	F160W	&	2025-04-24 14:53:02	&	1058.819	&	2020.7	&	F160W	&	16019	\\
            1RXS\,J1708(49.0-400910)	&	2025.3	&	F125W	&	2025-05-06 14:17:49	&	1058.819	&	2019.8	&	F125W	&	15348	\\
            	&	2025.3	&	F160W	&	2025-05-06 16:13:18	&	1058.819	&	2019.8	&	F160W	&	15348	\\
            CXOUJ1714(05.7-381031)	&	2025.3	&	F125W	&	2025-05-06 15:52:33	&	1058.819	&	2018.3	&	F125W	&	15348	\\
            	&	2025.3	&	F160W	&	2025-05-06 16:13:18	&	1058.819	&	2018.3	&	F160W	&	15348	\\
            SGR1900(+14)	&	2025.4	&	F125W	&	2025-05-26 15:38:49	&	1058.819	&	2020.7	&	F125W	&	16019	\\
            	&	2025.4	&	F160W	&	2025-05-26 15:59:34	&	1058.819	&	2020.7	&	F160W	&	16019	\\
            SGR1806(-20)$\dagger$	&	2025.4	&	F125W	&	2025-06-03 18:37:11	&	1058.819	&		&		&		\\
            	&	2025.4	&	F160W	&	2025-06-03 18:57:56	&	1058.819	&		&		&		\\
            AXJ1818(.8-1559)	&	2025.4	&	F125W	&	2025-06-04 11:59:48	&	1058.819	&	2018.6	&	F125W	&	15348	\\
            	&	2025.4	&	F160W	&	2025-06-04 12:20:33	&	1058.819	&	2018.6	&	F160W	&	15348	\\
            1E2259(+586)	&	2025.4	&	F125W	&	2025-06-15 01:29:31	&	1058.819	&	2018.6	&	F125W	&	15348	\\
            	&	2025.4	&	F160W	&	2025-06-15 01:50:16	&	1058.819	&	2018.6	&	F160W	&	15348	\\
            SGR1833(-0832)	&	2025.5	&	F125W	&	2025-06-16 06:56:38	&	1058.819	&	2018.5	&	F125W	&	15348	\\
            	&	2025.5	&	F160W	&	2025-06-16 08:30:27	&	1058.819	&	2018.5	&	F160W	&	15348	\\
            SGR1935(+2154)	&	2025.5	&	F125W	&	2025-06-30 16:54:07	&	1058.819	&	 	&		&	 	\\
            	&	2025.5	&	F160W	&	2025-06-30 17:14:52	&	1058.819	&	 	&		&	 	\\
            	&	 	&		&	 	&	 	&	2021.4	&	F140W	&	16505	\\
            	&	 	&		&	 	&	 	&	2016.4	&	F140W	&	14502	\\
            	&	 	&		&	 	&	 	&	2015.6	&	F140W	&	14055	\\
            	&	 	&		&	 	&	 	&	2015.2	&	F140W	&	14055	\\
            SGR0501(+4516)	&	2025.6	&	F125W	&	2025-08-14 01:39:26	&	1058.819	&	2020.6	&	F125W	&	16019	\\
            	&	2025.6	&	F160W	&	2025-08-14 02:00:11	&	1058.819	&	2020.6	&	F160W	&	16019	\\
            	&	 	&		&	 	&	 	&	2012.8	&	F160W	&	12672	\\
            	&	 	&		&	 	&	 	&	2010.8	&	F160W	&	12306	\\
            4U0142(+61)	&	2025.5	&	F125W	&	2025-06-29 01:59:10	&	1058.819	&	2018.0	&	F125W	&	15348	\\
            	&	2025.5	&	F160W	&	2025-06-29 02:19:55	&	1058.819	&	2018.0	&	F160W	&	15348	\\
            CXOUJ1647(10.2-455216)	&	2026.1	&	F125W	&	2026-02-04 23:15:52	&	1058.819	&	 	&		&	 	\\
            	&	2026.1	&	F160W	&	2026-02-04 23:29:42	&	1058.819	&	 	&		&	 	\\
            	&	 	&		&	 	&	 	&	2023.2	&	F115W (JWST)	&	1905	\\
            	&	 	&		&	 	&	 	&	2023.2	&	F150W (JWST)	&	1905	\\
            	&	 	&		&	 	&	 	&	2023.2	&	F200W (JWST)	&	1905	\\
            	&	 	&		&	 	&	 	&	2023.2	&	F277W (JWST)	&	1905	\\
            	&	 	&	 	&	 	&	    &   2023.2	&   F444W (JWST)	&	1905	\\
            	&	 	&		&	 	&	 	&	2018.4	&	F140W	&	14805	\\
            	&	 	&		&	 	&	 	&	2017.4	&	F140W	&	14805	\\
            \hline	
	\end{tabular}
    \newline
    $\dagger$ - Previous epochs for SGR1806 suffered from guide star acquisition failure, resulting in unusable data.
\end{table*}

\begin{figure*}
\centering
\includegraphics[width=0.99\textwidth]{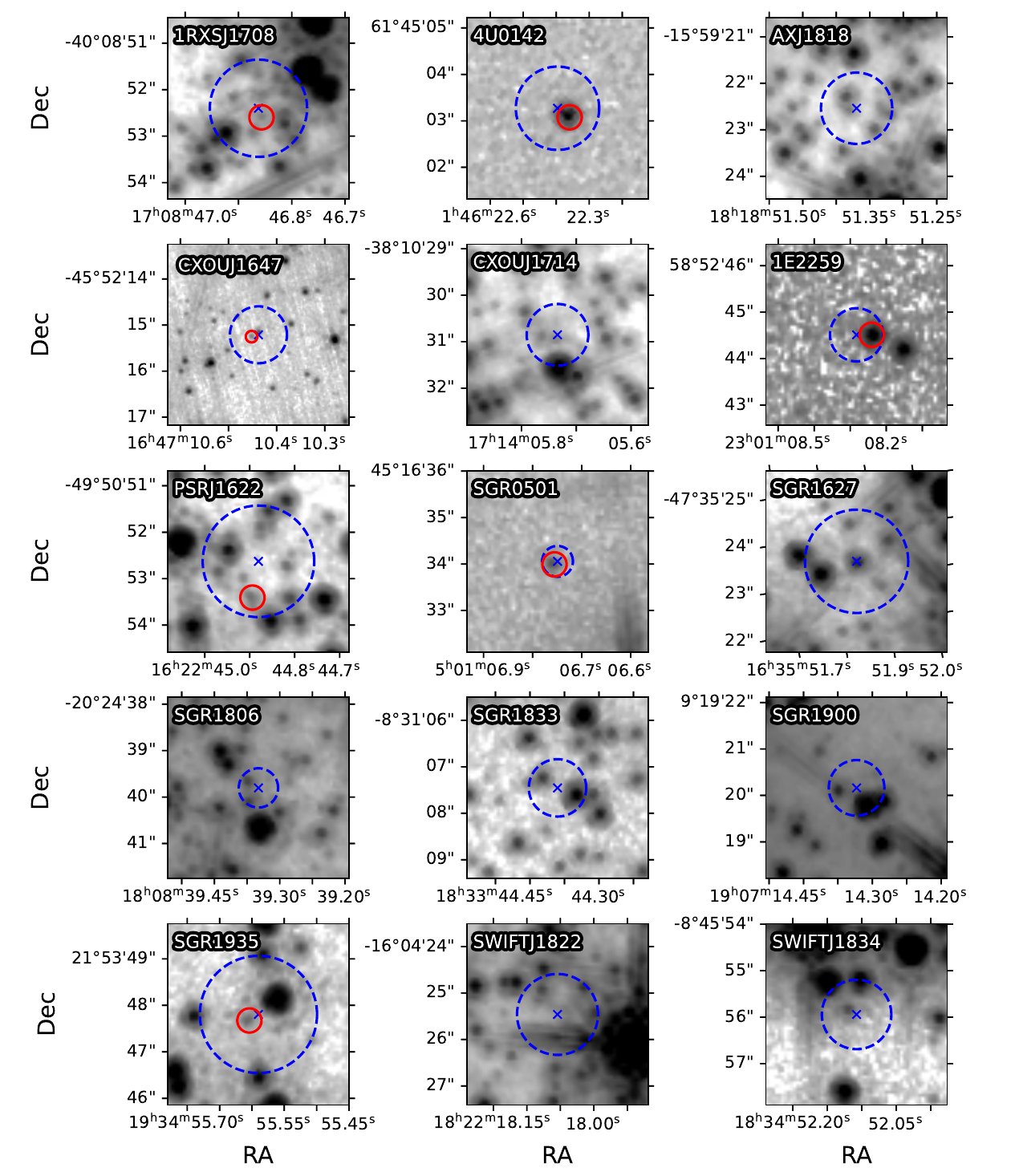}
\caption{HST cutouts with magnetar 3\,$\sigma$ X-ray localisations indicated by blue dashed circles with a cross at the centre \citep[from][and references therein]{2014ApJS..212....6O}. If a near-infrared counterpart candidate is identified, its position is indicated by a red circle. Only the latest (2025 epoch) F160W images are shown, with the exception of CXOUJ1647 (which is a 2023 JWST/NIRcam F150W image). Each cutout measures 4$\times$4\,arcsec, and they are oriented North up, East left.}
\label{fig:stamps}
\end{figure*}

\subsection{JWST}
Two objects in our sample have been imaged with JWST: 4U0142, as part of a campaign specifically designed to study the magnetar, and CXOUJ1647, whose location was covered by JWST imaging of the Westerlund 1 stellar cluster \citep{2025A&A...693A.120G}. A detailed JWST study of 4U0142 has already been published by \citet{2024ApJ...972..176H}. However, as the observations were taken with the SUB400P sub-array, the field of view is small and there are no reference sources we can use for astrometry (and hence a proper motion measurement) in the images. 

CXOUJ1647 on the other hand was imaged as part of a large mosaic covering the Westerlund 1 cluster (program GO 1905, PI: Guarcello). For this magnetar we have 2 previous epochs of F140W HST imaging, a 2026 epoch of F125W and F160W, plus imaging in 8 JWST wide filters (5 NIRcam and 3 MIRI, all obtained in 2023). We use the NIRcam/F150W image for the proper motion measurements, and make use of the {\sc dolphot} catalogue produced by \citet[][see next section]{2025A&A...693A.120G}. The data described here may be obtained from \url{https://dx.doi.org/10.17909/f9d0-er88}.


\subsection{Sample summary}
Our final sample consists of 14 objects with at least two epochs of space-based imaging. These are the 15 objects listed in Table \ref{tab:obsv}, minus SGR1806 whose 2025 epoch is the first set of HST images for this object. All remaining objects have two HST epochs (taken in 2018/2020 and 2025) with the exceptions of SGR0501 (three epochs of HST between 2010 and 2025), SGR1935 (five epochs of HST between 2015 and 2025) and CXOUJ1647 (four epochs, 3$\times$HST and 1$\times$JWST spanning 2017--2026).

In Figure \ref{fig:stamps} we show image cutouts (`stamps') around the position of each magnetar. The approximate 3$\sigma$ X-ray error circle is plotted in each case \citep[assuming Gaussian uncertainties and based on positions reported in the McGill magnetar catalogue,][we note that none of the 14 objects in our sample have a position measured with radio interferometry]{2014ApJS..212....6O}\footnote{\url{https://www.physics.mcgill.ca/~pulsar/magnetar/main.html}}. Sources which we identify as NIR counterparts to magnetars, based on the criteria outlined in the next section, are circled in red.

\section{Methodology}\label{sec:method}
\subsection{Photometry}
We carry out source detection and photometry for all HST images using {\sc dolphot} \citep{2000PASP..112.1383D,2016ascl.soft08013D,2024ApJS..271...47W}\footnote{\url{http://americano.dolphinsim.com/dolphot/}} with the point spread functions of \citet{2016wfc..rept...12A}. The detection thresholds for point sources were set at 3$\sigma$. Source detection was performed using only the individual {\sc \_flt} frames, with photometry of detected sources performed on both these and the drizzled {\sc \_drz} images. Magnitudes are converted to the AB system using conversions calculated with {\sc stsynphot}\footnote{\url{https://github.com/spacetelescope/stsynphot_refactor}}. These conversions (added to Vega magnitudes to obtain AB) are 0.9204 (F125W), 1.0973 (F140W), 1.2277 (F150W) and 1.2741 (F160W). For the JWST observations of CXOUJ1647 we use the existing {\sc dolphot} output catalogue of sources in and around Westerlund 1 produced by \citet{2025A&A...693A.120G}, as described in that paper. Additions to the JWST/NIRcam Vega magnitudes are 0.7949 (F115W), 1.2277(F150W), 1.7014 (F200W), 2.3160 (F277W) and 4.3664 (F444W) \citep{2020sea..confE.182R}\footnote{The SVO filter profile service}.

\subsection{Astrometry and proper motions}
By measuring the relative offset of a source in images taken some time apart, with respect to the surrounding objects in the field, we can measure its proper motion. We adopt a method described in detail by \citet{2018MNRAS.481.5339B} and \citet{2022ApJ...926..121L}, but briefly summarise it here. Starting from the {\sc dolphot} source catalogues, we first perform quality cuts \citep[removing objects which are not point sources or are overly crowded, as outlined by][]{2022ApJ...926..121L}. {\em Gaia} sources within the HST/WFC3 field of view are then selected \citep[using {\em Gaia} data release 3,][]{2023A&A...674A...1G}, and sources with a poor astrometric solution \citep[a renomalised unit weight error, or RUWE $>$1.2,][]{2022MNRAS.513.2437P} are removed. The {\em Gaia} positions were then offset, using their proper motions where available, to obtain their expected positions at the time of each HST observation. These shifted positions (in equatorial coordinates) are then transformed first into a tangent plane, and then to HST pixel (x,y) coordinates using the existing world coordinate system of the images (which itself has been determined through an astrometric tie with {\em Gaia}). The x,y coordinates of the shifted {\em Gaia} sources are then matched with the HST source catalogues by identifying the closest match to each {\em Gaia} source. Outliers in the resulting offset distribution for the matching pairs (above the 95$^{\rm th}$ percentile) are rejected. A new (x,y) to ($\alpha$cos($\delta$), $\delta$) transformation is then fit following \citet{2018MNRAS.481.5339B}, where the uncertainty in this transformation represents the absolute astrometric uncertainty.

\begin{figure*}[h!]
\centering
\includegraphics[width=0.99\textwidth]{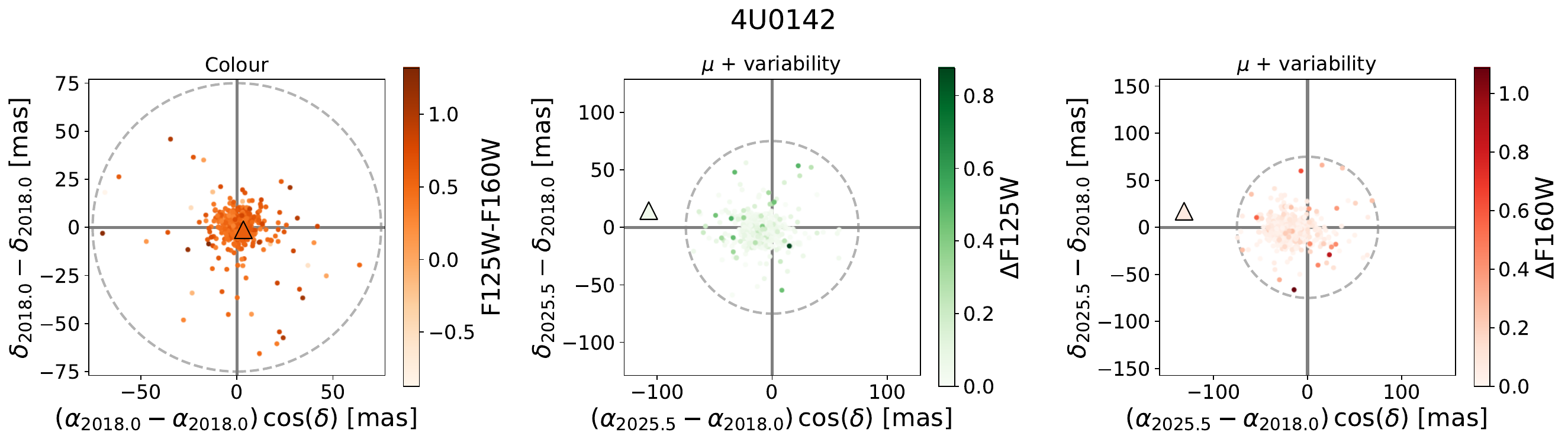}
\includegraphics[width=0.99\textwidth]{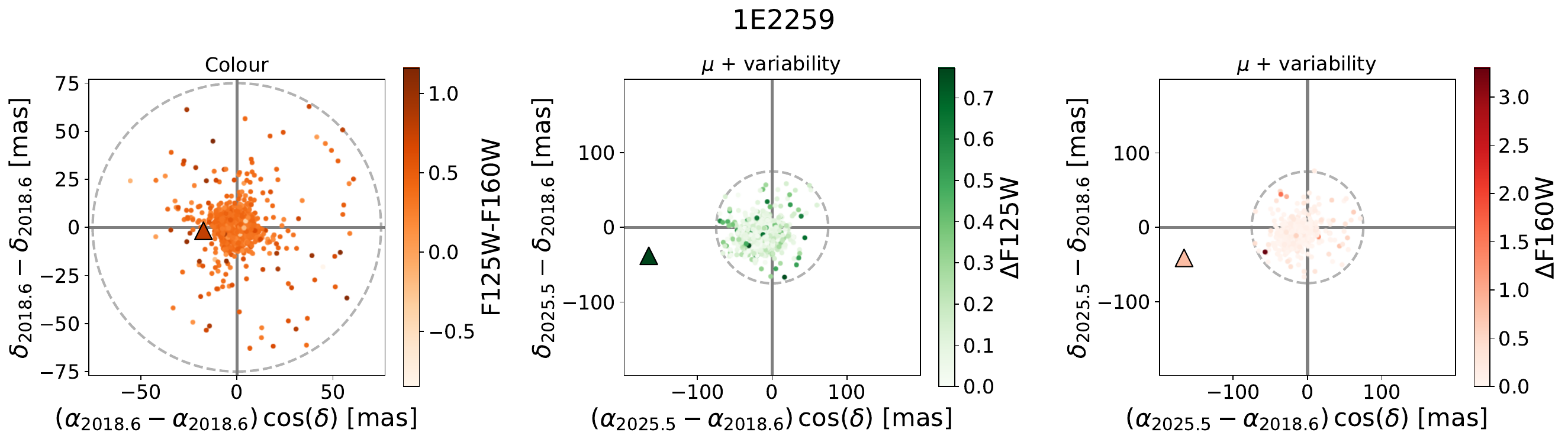}
\includegraphics[width=0.99\textwidth]{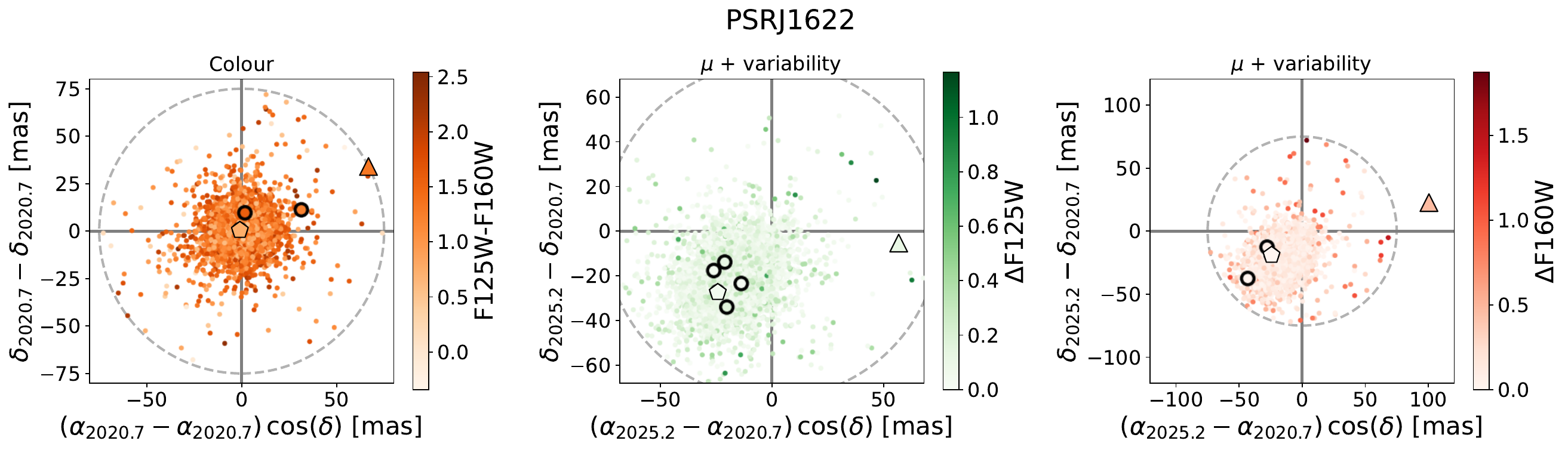}
\includegraphics[width=0.99\textwidth]{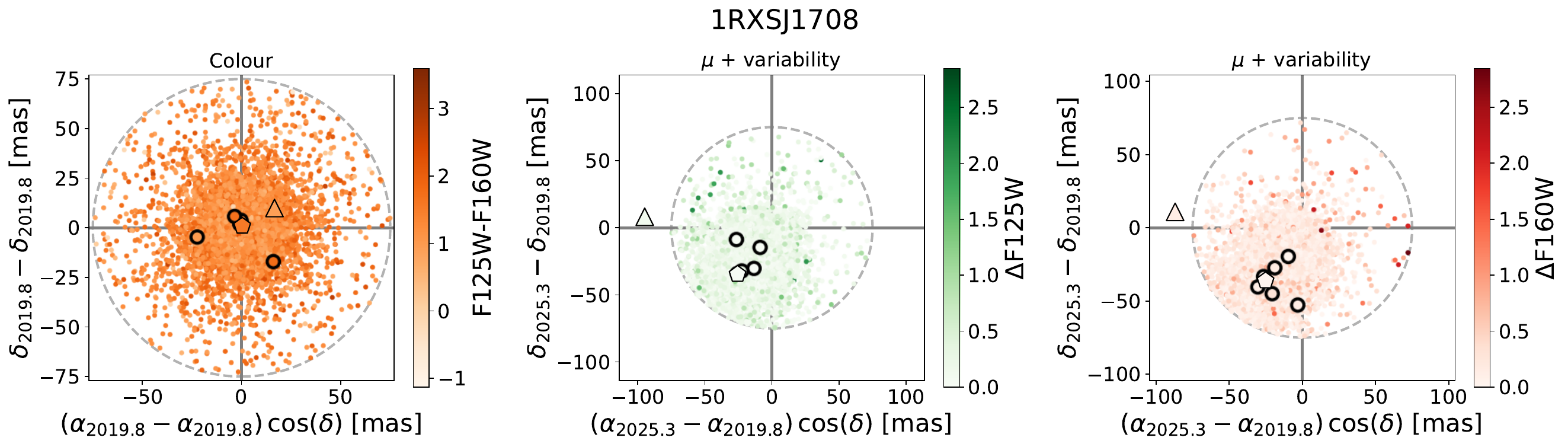}
\caption{Figures showing the change in RA and Dec of sources in the HST images between epochs. Shown here are the cases where a good counterpart candidate was identified in the 3$\sigma$ X-ray error circles shown in Figure \ref{fig:stamps}, based on distinct kinematics, colour and/or variability, and four images were available (2 filters in 2 epochs). In each case, our preferred counterpart candidate is labelled with a triangle. The best candidate in the 3$\sigma$ X-ray error region per the P$_{\rm chance}$ criteria is marked with pentagon \citep[if different from our favoured counterpart, see also][]{2022MNRAS.512.6093C}. The lowest P$_{\rm chance}$ values in each case are (top to bottom) $<$10$^{-3}$, 0.016, 0.043 and 0.164. Other objects in the X-ray localisation region are labelled as circles with bold outlines, with the colouring of all markers following the same mapping as the cloud of field sources. Left column: comparing the positions of sources in images taken minutes apart, showing astrometric noise. The points correspond to objects detected in both HST images and are coloured according to their F125W-F160W colour. Middle: the movement of sources between the first and second epochs of F125W imaging. The bulk motion of sources in the field is clear, magnetars tend to stand apart from this systemic drift due to their natal kicks. The points are coloured by the difference in F125W between the epochs. Right: as for the middle panels, but showing the difference between the two epochs in F160W. The grey dashed circles represent the matching radius (75\,mas) used to identify common sources between each pair of images.}
\label{fig:cp_ident_yes}
\end{figure*}

This method yields absolute astrometry at a comparable level to HST-only relative astrometry \citep[typically several tens of milliarcseconds,]{2022ApJ...926..121L}. The uncertainty on the resulting proper motion of a given source has contributions from both the absolute astrometric uncertainty and the positional uncertainty of the source (i.e. how well its centroid can be measured, which is proportional to the signal-to-noise). Because we have {\sc dolphot} catalogues containing positions for all sources in the images above a detection threshold, we can plot, for any pair of images, the difference ($\Delta \alpha$cos($\delta$), $\Delta \delta$) between sources which are detected in both images. This is shown in Figures \ref{fig:cp_ident_yes},  \ref{fig:cp_ident_yes2},  \ref{fig:cp_ident_no},  \ref{fig:cp_ident_no2}. Sometimes we are comparing the positions of objects in images taken just minutes apart (e.g. two images with different filters in the same HST orbit). In this case, the scatter is centred on 0,0 and simply reflects the astrometric uncertainties. When the images are taken several years apart, the bulk motion of stars in the field is evident as a shift in the clouds of points. This apparent bulk motion can arise from both the differential rotation between our position in the Galaxy and along the sight-line in question; and the `real' motion of stars in their local standard of rest (e.g. moving groups). When reporting proper motions in Section \ref{sec:results}, we first account for the differential rotation \citep[and Solar velocity with respect to the local standard of rest, LSR,][]{2019RAA....19...68D} following \citet{2017A&A...608A..57V,2022ApJ...926..121L}. The reported proper motions are therefore proper motions with respect to the LSR at the assumed distance of the magnetar (see Table \ref{tab:mu}), i.e., they are peculiar proper motions. The impact of the large distance and hence velocity uncertainties are explored in Section \ref{sec:discuss}).

\afterpage{\clearpage}

\subsection{Counterpart identification}\label{sec:ident}
\begin{figure*}[h!]
\centering
\includegraphics[width=0.9\textwidth]{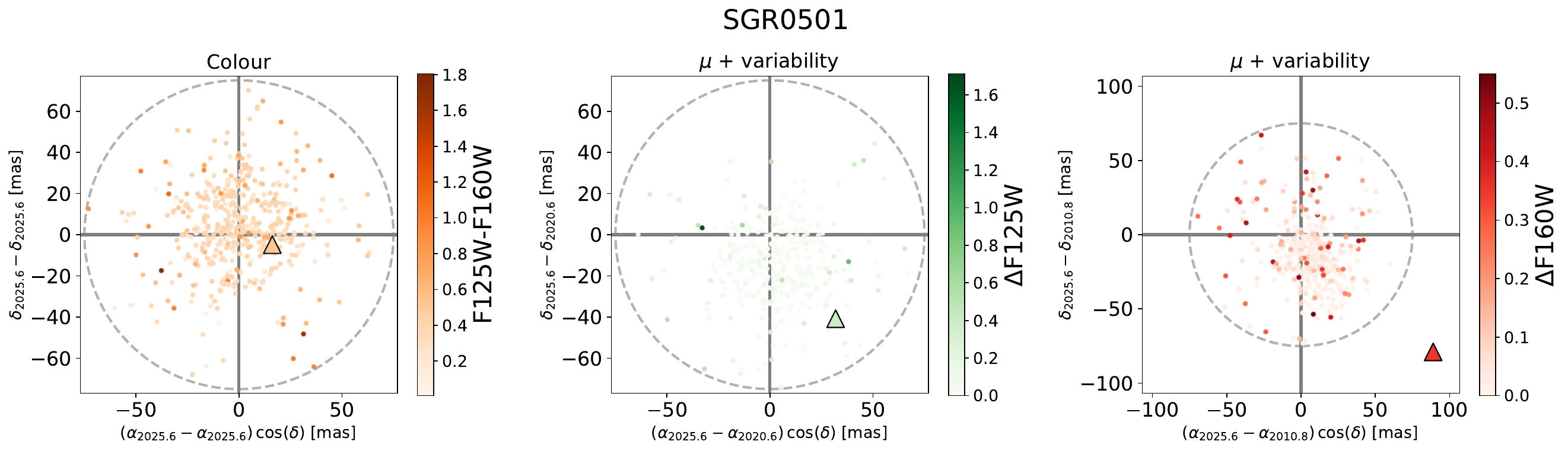}
\includegraphics[width=0.9\textwidth]{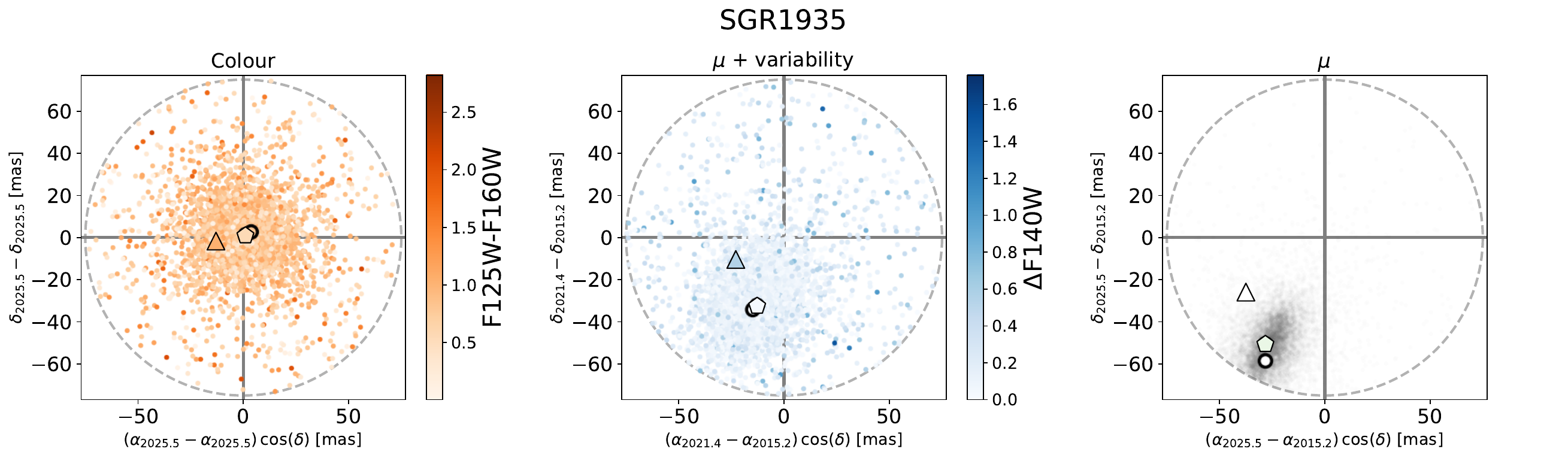}
\caption{As in Figure \ref{fig:cp_ident_yes}, but for the cases where six images were available. For each filter we only show comparisons for the greatest temporal difference (e.g. 2010 versus 2025 for the F160W observations of SGR0501). The $\mu$-only panel for SGR1935 has no colourbar as it compares different filters at different times (F140W versus F125W), but is included nevertheless as it represents the greatest temporal baseline (and hence greatest change in RA and Dec) available for this magnetar. The lowest P$_{\rm chance}$ values are 0.002 (SGR0501) and 0.02 (SGR1935, this is the adjacent star, not the counterpart).}
\label{fig:cp_ident_yes2}
\end{figure*}

\begin{figure*}[h!]
\centering
\includegraphics[width=0.9\textwidth]{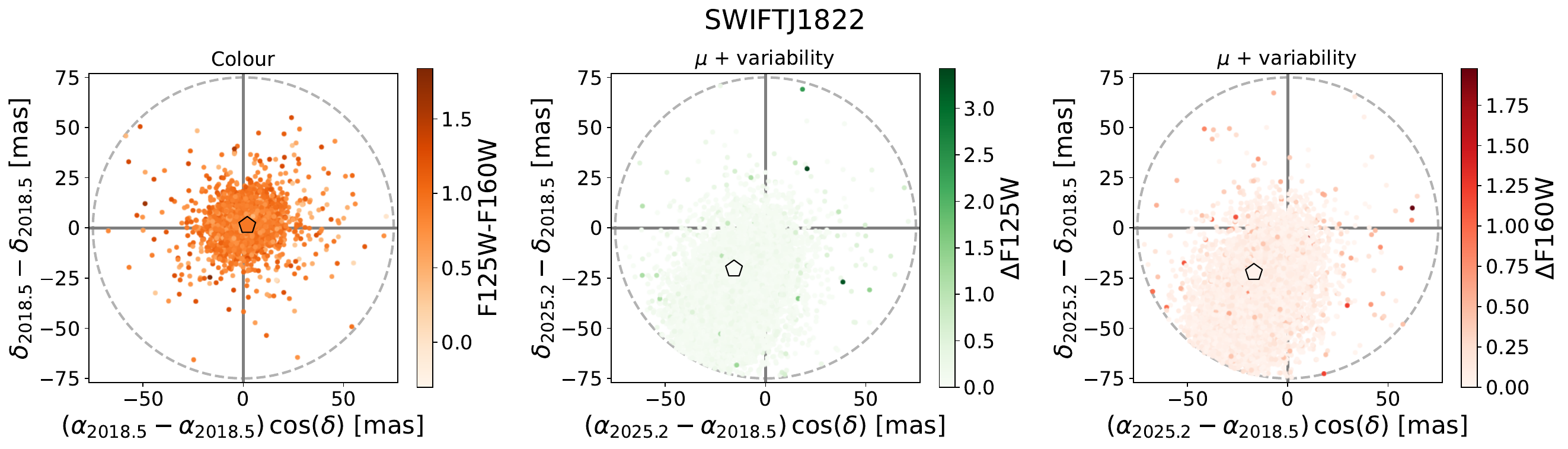}
\includegraphics[width=0.9\textwidth]{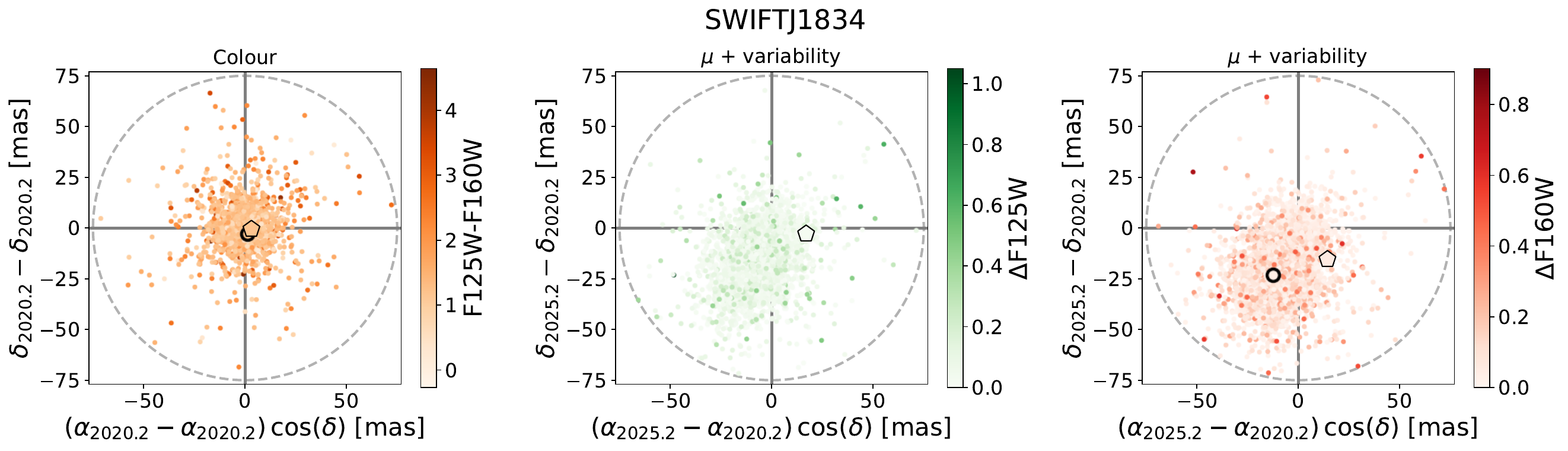}
\includegraphics[width=0.9\textwidth]{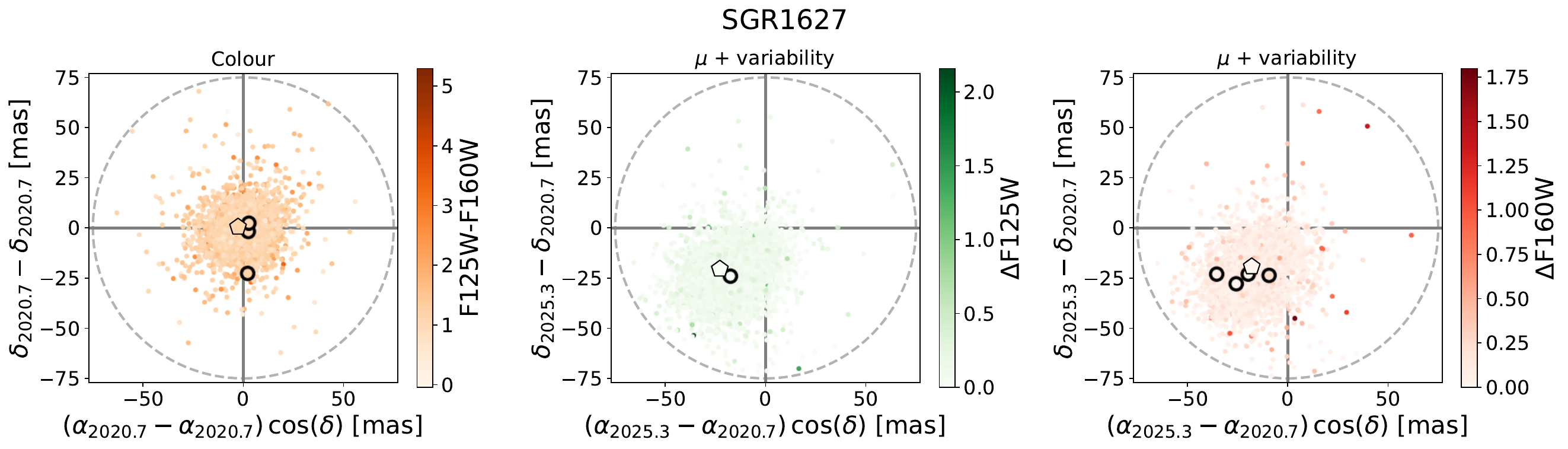}
\includegraphics[width=0.9\textwidth]{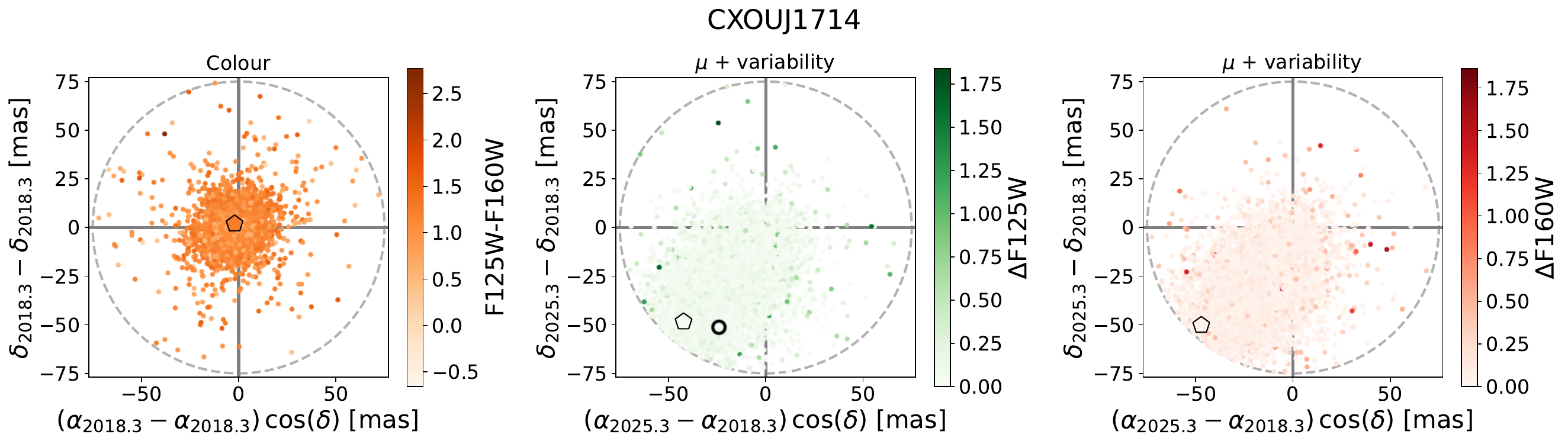}
\caption{As in Figures \ref{fig:cp_ident_yes} and \ref{fig:cp_ident_yes2}, but for the cases where no good candidate was identified in the 3$\sigma$ error circle based on distinct kinematics, colour and/or variability. The sources marked with pentagons are the candidates selected solely on P$_{\rm chance}$ grounds by \citet{2022MNRAS.512.6093C}, the P$_{\rm chance}$ values are (top to bottom) 0.15, 0.05, 0.03 and 0.005.}
\label{fig:cp_ident_no}
\end{figure*}

\begin{figure*}[h!]
\centering
\includegraphics[width=0.9\textwidth]{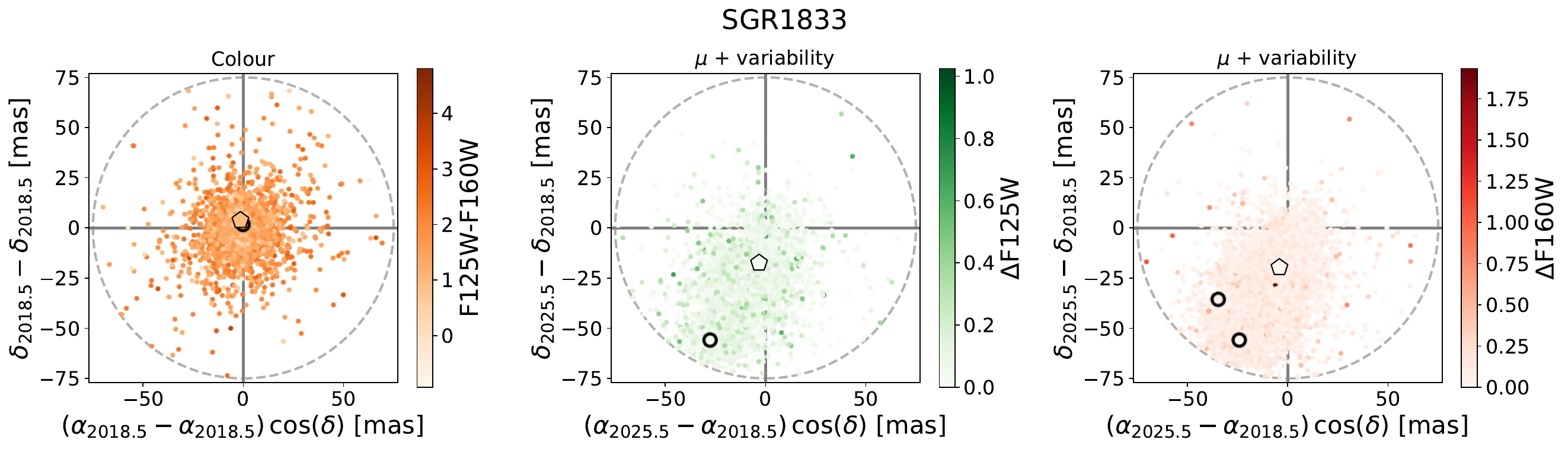}
\includegraphics[width=0.9\textwidth]{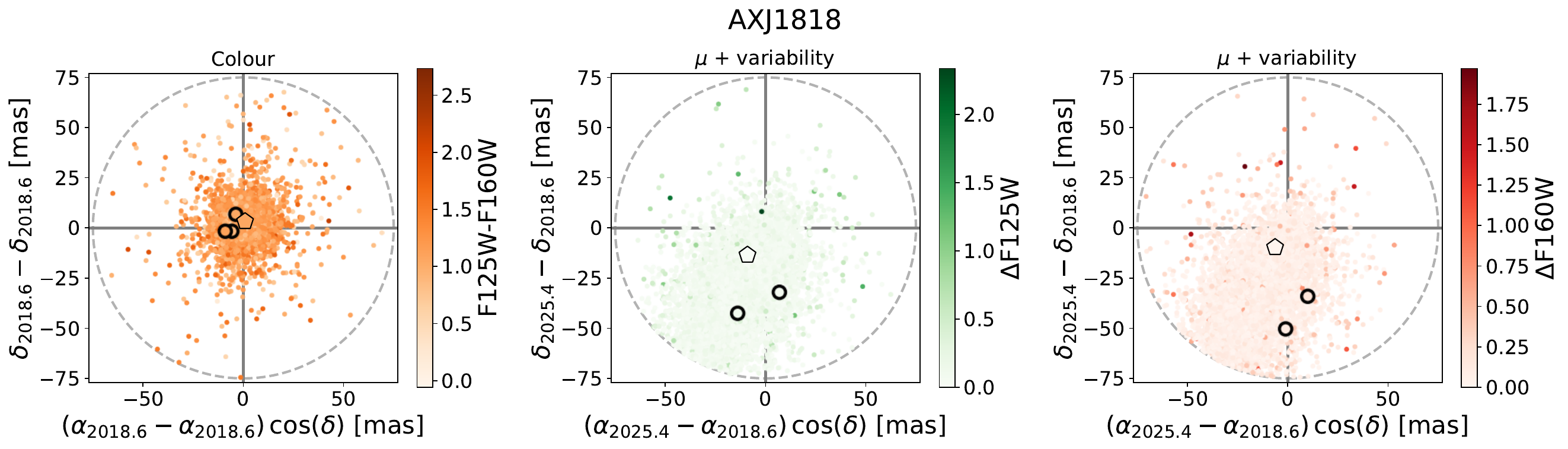}
\includegraphics[width=0.9\textwidth]{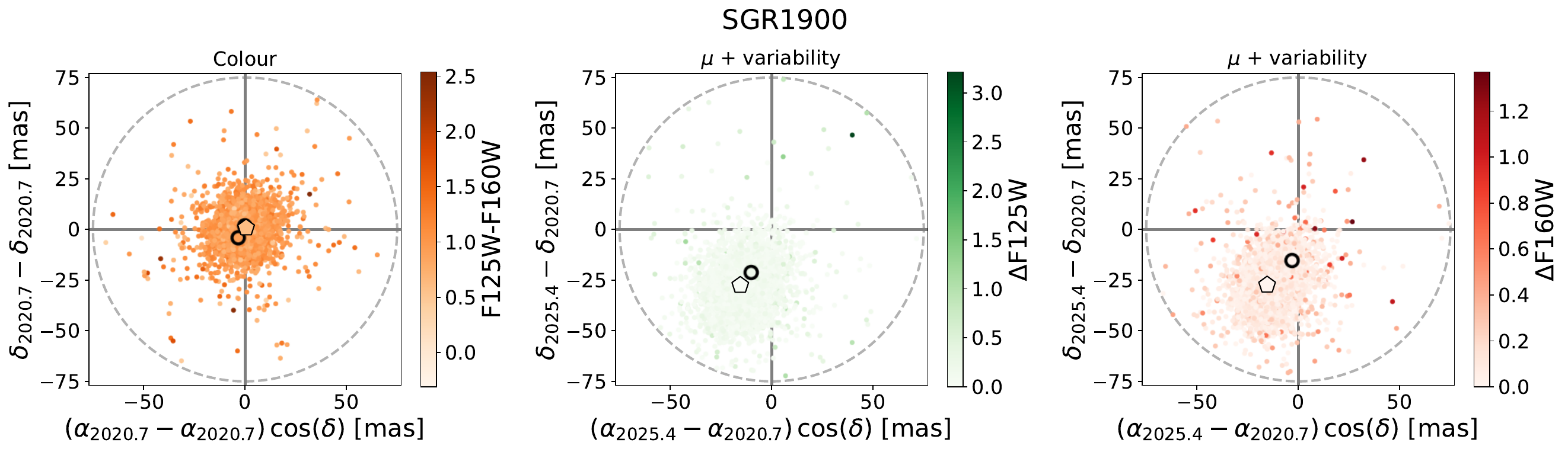}
\includegraphics[width=0.9\textwidth]{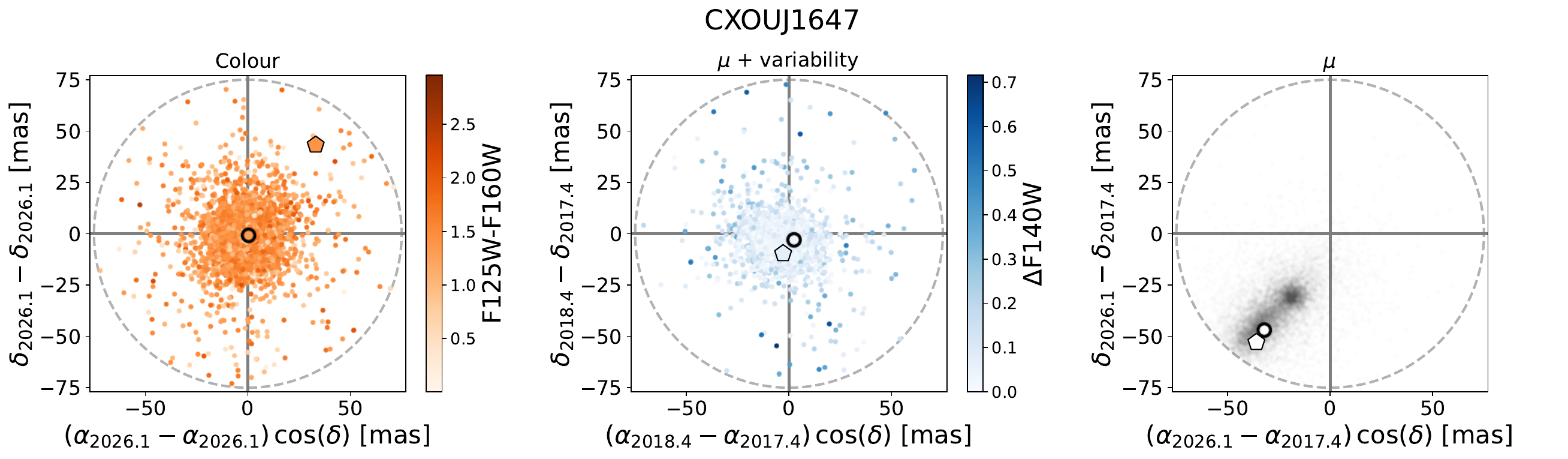}
\caption{Figure \ref{fig:cp_ident_no} continued. More cases where no good candidate was identified in the 3$\sigma$ error circle based on distinct kinematics, colour and/or variability. The sources marked with pentagons are simply the candidates identified on P$_{\rm chance}$ grounds by \citet{2022MNRAS.512.6093C}, the P$_{\rm chance}$ values are (top to bottom) 0.03, 0.18, 0.03 and 0.45. The last panel for CXOUJ1647 has no colourbar because it compares the first epoch in 2017 (F140W) with the JWST/NIRcam F150W image taken in 2023. Our preferred counterpart for CXOUJ1647 is not detected in the HST images and therefore does not appear on these plots. }
\label{fig:cp_ident_no2}
\end{figure*}

To identify NIR counterparts, we follow an established methodology of searching for objects in the magnetar X-ray error circles which are discrepant from objects in the field in terms of proper motion, colour and variability. This is particularly useful in crowded fields with multiple candidates, and because magnetars are faint (and potentially undetected) at these wavelengths.
In most cases we have 2 HST epochs, with F125W and F160W imaging in each. In Figures \ref{fig:cp_ident_yes}, \ref{fig:cp_ident_yes2}, \ref{fig:cp_ident_no} and \ref{fig:cp_ident_no2} we plot the positional shifts ($\Delta \alpha$cos($\delta$), $\Delta \delta$) for all NIR sources in the field which are matched between the images. A matching radius of 75\,mas is used, which is indicated in these figures by dashed grey circles. This radius is extended for objects inside the X-ray error circles to enable the identification of high proper motion objects, this point is discussed in more detail in Section \ref{sec:discuss}. In the standard case we have one panel where the two images are taken minutes apart (same epoch, different filters). In these panels, the colourmap represents the F125W-F160W colour, and the $\Delta \alpha$cos($\delta$), $\Delta \delta$ scatter solely reflects astrometric noise. We also have panels showing $\Delta \alpha$cos($\delta$), $\Delta \delta$ for pairs of F125W and F160W images, obtained years apart. Here the colour scale represents variability instead (i.e. $\Delta$ F125W, $\Delta$ F160W), and the distribution in $\Delta \alpha$cos($\delta$), $\Delta \delta$ parameter space has contributions from both astrometric noise and proper motion as described in the previous section. 

In every panel of these figures, objects inside the 3$\sigma$ X-ray error circles are indicated by circles with bold outlines, and the best counterpart candidate based on proper motion, colour and variability is marked with a triangle. To classify an object as a counterpart candidate, we require either (i) at least one of the proper motion, colour or variability to be $>$3$\sigma$ outside the scatter in the field objects, in at least one epoch or (ii) two or more of proper motion, colour or variability to be $>$2$\sigma$ outliers. Figures \ref{fig:cp_ident_yes} and \ref{fig:cp_ident_yes2} show the magnetars where a kinematically distinct, red and/or variable object was found in the X-ray error circle. If different from this candidate, the highest probability source based on the probability of chance alignment \citep[P$_{\rm chance}$, e.g.][]{2022MNRAS.512.6093C} is marked with a pentagon. We note that in each case where there is more than one candidate in the error circle, our preferred candidate is not the one with the lowest P$_{\rm chance}$. This demonstrates that P$_{\rm chance}$ is not an effective selection method for magnetar counterparts, perhaps not surprising given that their faint magnitudes, and the nature of the P$_{\rm chance}$ selection which favours brighter objects. All markers follow the same colour-mapping as the cloud of field sources. 

Six objects in our sample meet one of the proper motion, colour or variability criteria: 4U0142 and 1RXSJ\,1708 pass purely on their discrepant proper motion, SGR1935 passes on variability alone,  SGR0501 and PSRJ1622 passes on both proper motion and variability, and 1E2259 passes on all three criteria. In every case the parameter on which the selection is made exceeds 3$\sigma$. Of these, the counterparts of 4U0142, 1E2259, SGR0501, SGR1935 were already known, further supporting the validity of the method. We additionally report new NIR counterparts for PSRJ1622 and 1RXSJ\,1708.  

4U0142, 1E2259, PSRJ1622 and 1RXSJ\,1708 have just two epochs of two-filter HST imaging. SGR0501 has four epochs of observation, the first two solely in F160W and last two in both F125W and F160W. Using the 2025 epoch as the reference we therefore have three $\Delta F160W$, one $\Delta F125W$ and one F125W-F160W as the possible combinations. For brevity in Figure \ref{fig:cp_ident_yes2} we only show F125W-F160W, $\Delta$F125W and the $\Delta$F160W with the longest temporal baseline. The known counterpart of SGR0501 clearly stands out in terms of kinematics and variability. SGR1935 has five epochs: four in which the sole filter was F140W, and 2025 with F125W and F160W. In Figure \ref{fig:cp_ident_yes2} we show the 2025 F125W-F160W colour, the $\Delta$F140W with the longest baseline and finally an F125W (2026) versus F140W (2017) comparison. The last panel represents the largest temporal baseline available for this object, but we cannot disentangle colour/variability when comparing different filters taken years apart, so no colour-mapping is applied. The proper motion of SGR1935 is clearly visible, particularly in this final panel.

The rest of the sample is shown in Figures \ref{fig:cp_ident_no} and \ref{fig:cp_ident_no2}. In these cases, none of the sources in the X-ray localisation meet the selection criteria, so we cannot confidently claim to have identified the counterpart. The `best candidate' in Figures \ref{fig:cp_ident_no} and \ref{fig:cp_ident_no2} is labelled as such purely on the chance alignment probability (i.e. they are the brightest sources in the error circle).

\section{Results}\label{sec:results}\label{sec:cps}
    We present the apparent magnitudes, fluxes, colours and variability of confirmed/preferred counterparts in Table \ref{tab:mags}, and LSR-corrected proper motions for the six counterparts in Table \ref{tab:mu}. The latter also includes distance estimates for the magnetars, and the resulting peculiar transverse velocities $v_{\rm t}$ at that distance. Notes on individual objects for which we report a new counterpart candidate, fail to recover a previously known counterpart, or successfully `re-discover' a known source, are provided in the following subsections. 
    
    These results bring the total number of magnetar IR counterparts to 11 (two of which are newly reported here) out of $\sim$30 known \citep{2026enap....3..205R}. Those with IR counterparts are the six shown in Figures \ref{fig:cp_ident_yes} and \ref{fig:cp_ident_yes2}, plus SGRs 1806 and 1900 \citep{2005ApJ...623L.125K,2012ApJ...761...76T}, 1E1048.1-5937 \citep{2008ApJ...677..503T}, XTEJ1810-197 and 1E1841-045 \citep{2008A&A...482..607T}

\subsection{Recovered known counterparts}
We successfully recover the known NIR counterparts of 1E2259 \citep{2001ApJ...563L..49H,2004ApJ...617L..53T,2013ApJ...772...31T}, 4U0142 \citep{2004A&A...416.1037H,2006Natur.440..772W,2024ApJ...972..176H}, SGR1935 \citep{2018ApJ...854..161L,2022ApJ...926..121L} and SGR0501 \citep{2011MNRAS.416L..16D,2025AandA...696A.127C}. These objects all have a previously measured proper motion, which are provided for reference in Table \ref{tab:mu}, and we reduce the uncertainties further by moving to space-based observations and/or extending the temporal baseline.

\subsection{Unrecovered known counterparts}
\subsubsection{SGR1806-20}
The 2025 epoch imaging presented in this paper is the first space-based imaging of SGR1806, although a magnitude 23--34, variable K$_{\rm s}$-band counterpart was identified by \citet{2005A&A...438L...1I,2005ApJ...623L.125K} and subsequently in ground-based Very Large Telescope (VLT) and Keck imaging \citet{2012ApJ...761...76T}. There is no detected source at the position of the source labelled A by \citet{2005A&A...438L...1I,2012ApJ...761...76T} in the 2025 F125W and F160W images. There is however a slightly offset source present in the F160W images (the closest in terms of wavelength) which lies between the position of SGR1806 in 2004 and the star to the south-west which \citet{2005A&A...438L...1I,2012ApJ...761...76T} labelled C. Assuming this is SGR1806 and comparing its position with the reported 2004 coordinates, it would have a proper motion of $\sim$10\,mas\,yr$^{-1}$, higher than the value of $\sim$5\,mas\,yr$^{-1}$ reported by \citet[][]{2012ApJ...761...76T} and in an approximately perpendicular direction. We therefore assume that the F160W source is not related, and that SGR1806 is not detected in our F125W and F160W imaging. The high extinction along the sight-line means that our imaging, even in the longer F160W filter, could reduce the counterpart flux as much as 6 magnitudes \citep[for a flat spectrum and A$_{V}\sim30$][]{2004ApJ...616..506E} versus the 23--34 K$_{\rm s}$-band magnitude, plausibly explaining our non-detection.

\subsubsection{SGR1900$+$14}\label{sec:1900}
The infrared counterpart of SGR1900 was previously identified in VLT/NACO \citep{2008A&A...482..607T} and Keck/NIRC2 \citep{2012ApJ...761...76T} K$_{\rm s}$-band imaging, using adaptive optics for enhanced spatial resolution which reached the $\sim$ milli-arcsecond level. Comparing the images in those papers with Figure \ref{fig:stamps}, we can see that the source is not detected in our imaging. This is likely due to the factor of $\sim$ 10 lower spatial resolution and subsequent blending with the adjacent star, rather than variability, since our HST imaging has a limiting magnitude which is 5-6 magnitudes deeper. Higher resolution observations with current adaptive optics-assisted ground-based facilities, JWST, or future observatories such as the the European Extremely Large Telescope \citep{2007Msngr.127...11G} would likely be required to re-detect the counterpart. In any case, this magnetar already has a proper motion measurement \citep[][see Table \ref{tab:mu}]{2012ApJ...761...76T}.

\subsection{New counterpart candidates}
\subsubsection{PSRJ1622$-$4950}\label{sec:1622}
We report the discovery of a candidate infrared counterpart to PSRJ1622$-$4950. The object is one of four detected in the 3$\sigma$ X-ray error circle. It is kinematically distinct with a high proper motion, and shows variability (most notably in F160W). This is not the candidate identified by \citet{2022MNRAS.512.6093C}, which was simply selected based on P$_{\rm chance}$ arguments. 

\citet{2010ApJ...721L..33L} report a distance for PSRJ1622$-$4950 (and the associated SNR\,G333.9+0.0) of $\sim$9\,kpc (with an uncertainty of a factor of $\sim$2), based solely on the radio dispersion measure using the NE2001 electron density model \citep[][]{2002astro.ph..7156C}. Placing the IR counterpart of PSRJ1622$-$4950 at 9\,kpc, and correcting for the extinction at 1.6$\mu$m of 5.2 magnitudes 
\citep[adopting a Fitzpatrick extinction law,][and R$_{\rm V}=3.1$]{1999PASP..111...63F}, we find the source is extremely blue, with an intrinsic absolute magnitude of M$_{\rm F160W}\sim3.5$. This is $\sim$5--8 magnitudes brighter than other confirmed IR counterparts \citep[e.g.][]{2022MNRAS.513.3550C}. Placing it instead at 5.5\,kpc, the distance implied by the YMW16 electron density model \citep{2017ApJ...835...29Y}, and adopting approximately half the extinction as implied from the A$_{\rm V}$-N${\rm H}$ relation \citep{1995A&A...293..889P}, we find M$_{\rm F160W}\sim8$ which is more comparable with other counterparts. Overall, a distance of $\sim$5\,kpc brings PSRJ1622$-$4950 into line with the rest of the population, and given the large uncertainty on the previous distance estimate, such a revision is plausible. We therefore suggest than this object (and its associated SNR) lie at $\sim$5\,kpc, adopting this revised distance throughout. Given the NIR luminosity this implies ($\sim 5 \times 10^{31}$erg\,s$^{-1}$), the IR efficiency for this magnetar, defined as log$_{10}$(L$_{\rm IR}$/$\dot{E}$) \citep{2010MNRAS.407.1887R}, is -2.2, where we have adopted $\dot{E} = 8.5 \times 10^{33}$\,erg\,s$^{-1}$ \citep{2010ApJ...721L..33L}. Previous measurements for magnetars have yielded values in range -4 to -1. 

The position of the PSRJ1622 counterpart candidate shows a systematic offset between the F125W and F160W filters at the same epoch (see Figure \ref{fig:pmfit1}). This may be due to a faint, adjacent source which has affected the centroiding (and possibly also the flux measurement). However, the proper motion appears to be consistent between the two filters despite this offset, and is small nevertheless, so this does not affect our conclusion regarding the origin of PSRJ1622 in SNR\,G333.9+0.0.

\begin{figure*}
\centering
\includegraphics[width=0.33\textwidth]{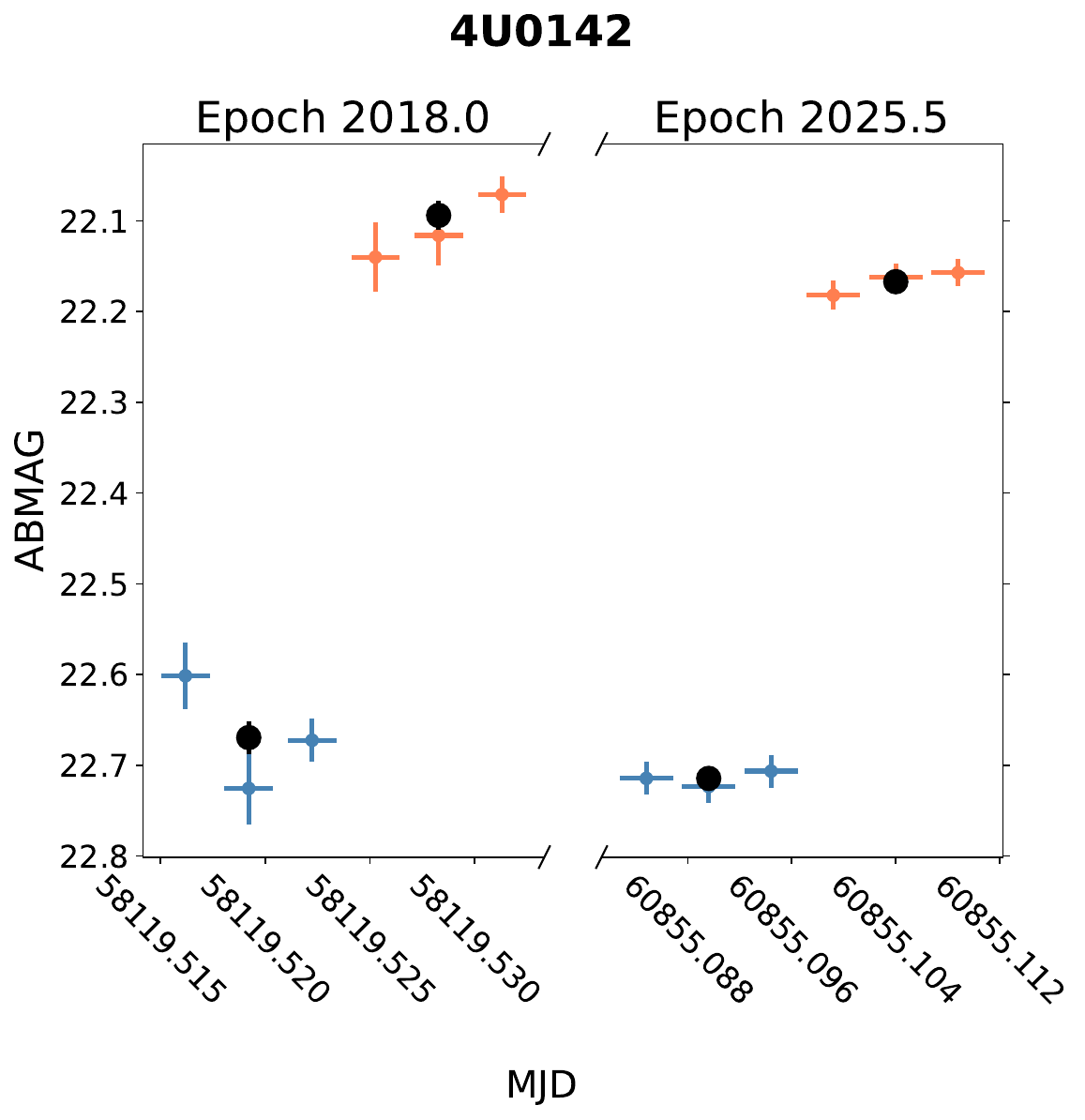}
\includegraphics[width=0.33\textwidth]{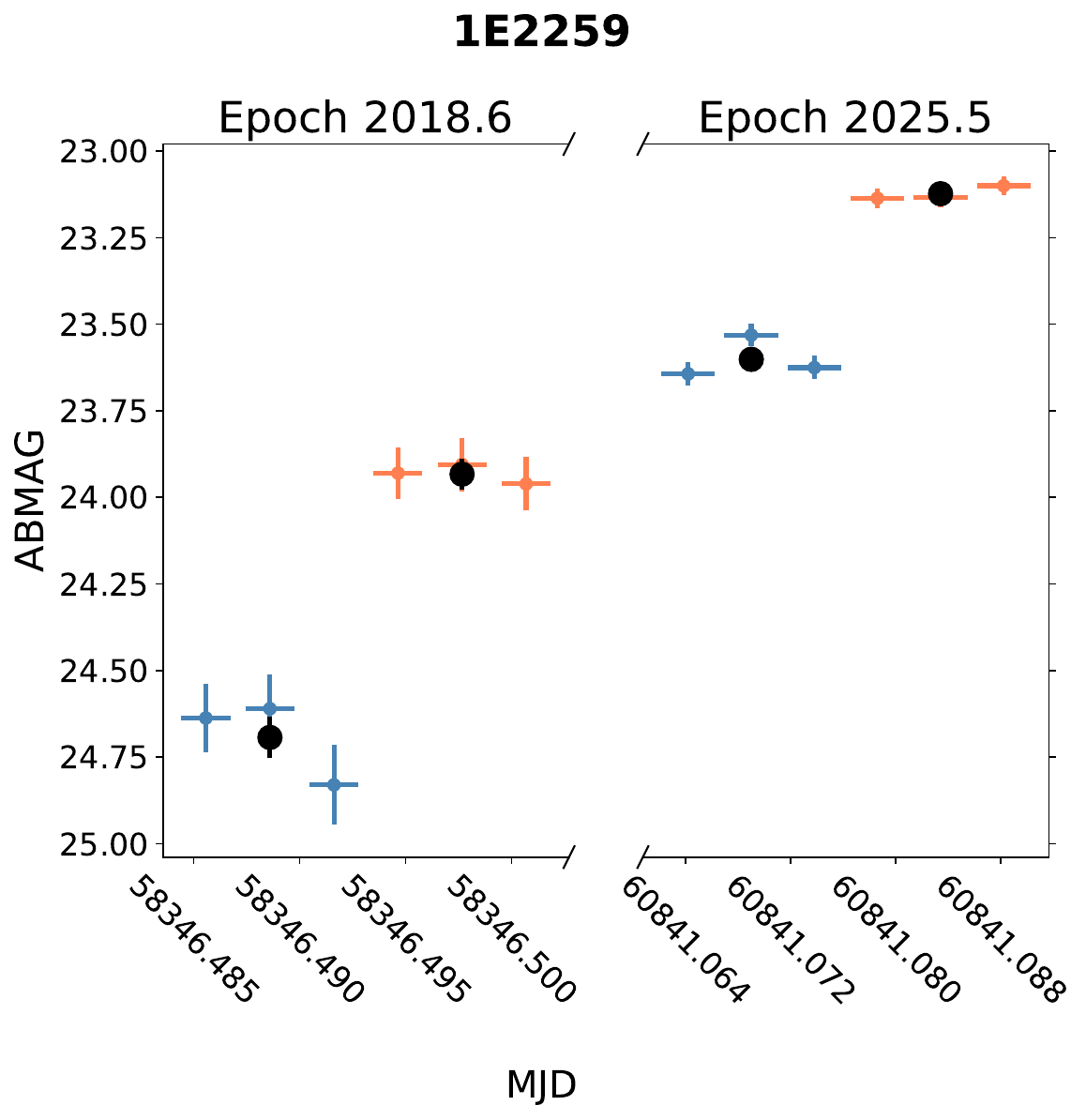}
\includegraphics[width=0.33\textwidth]{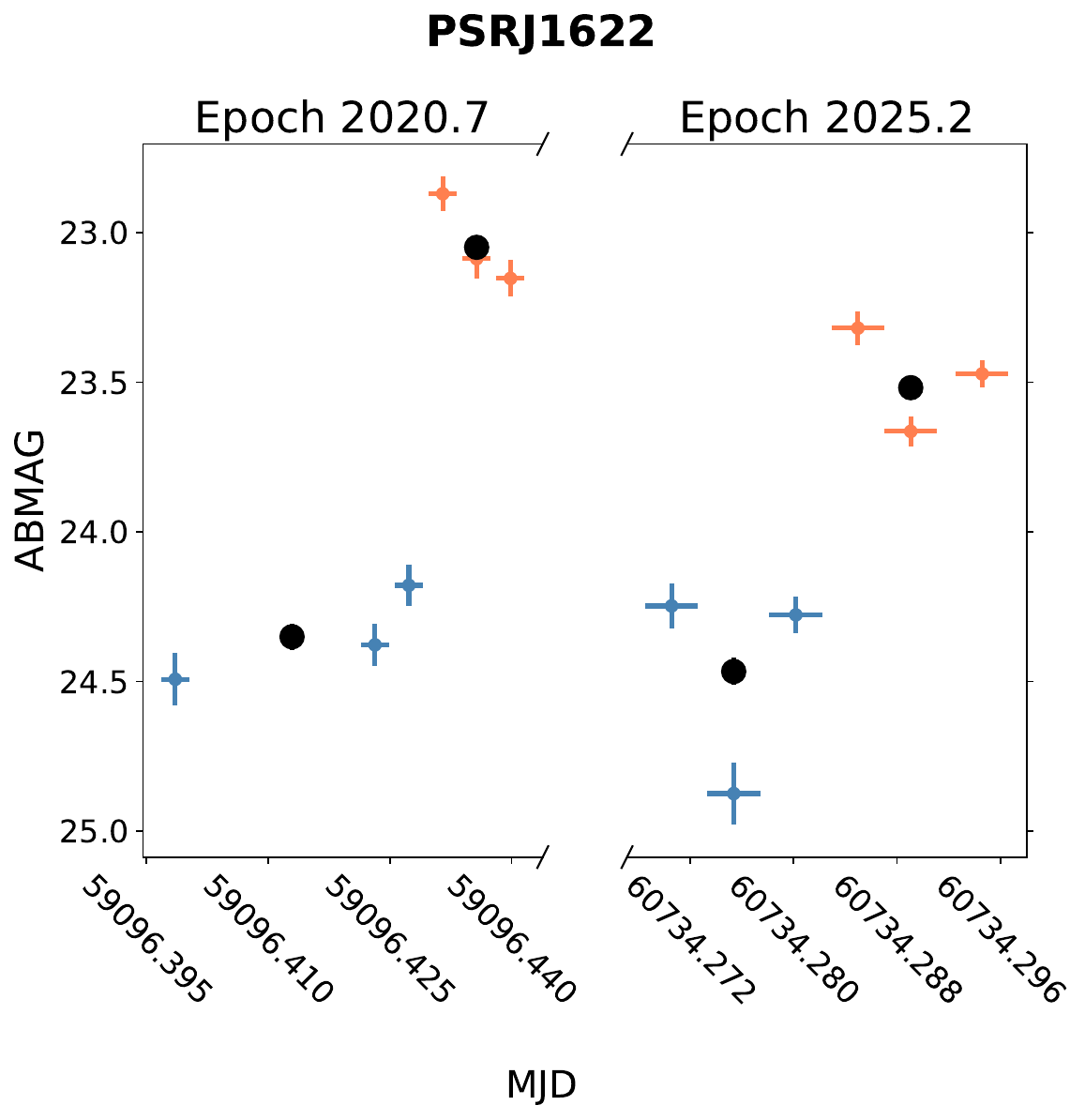}
\includegraphics[width=0.63\textwidth]{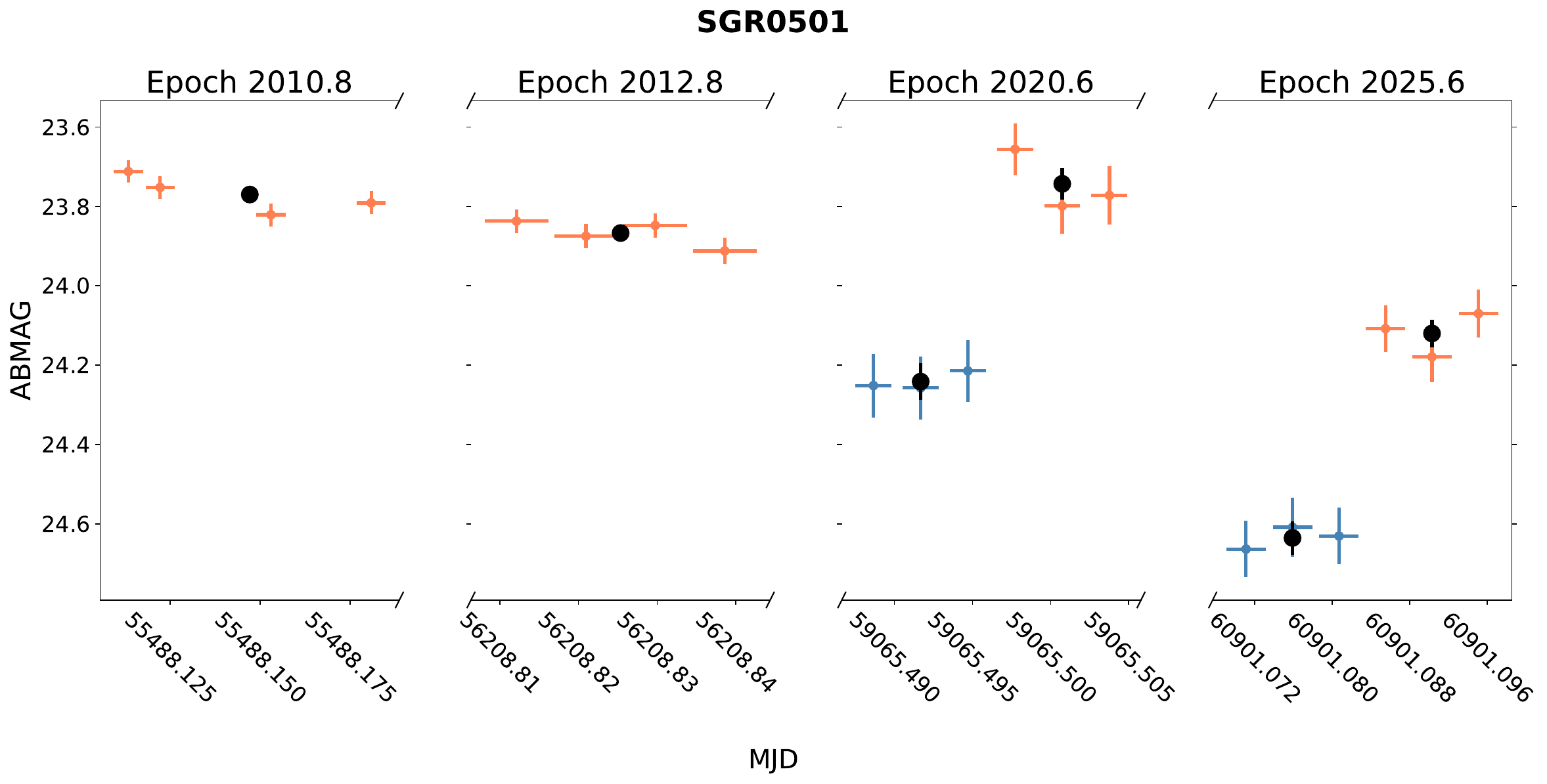}
\includegraphics[width=0.32\textwidth]{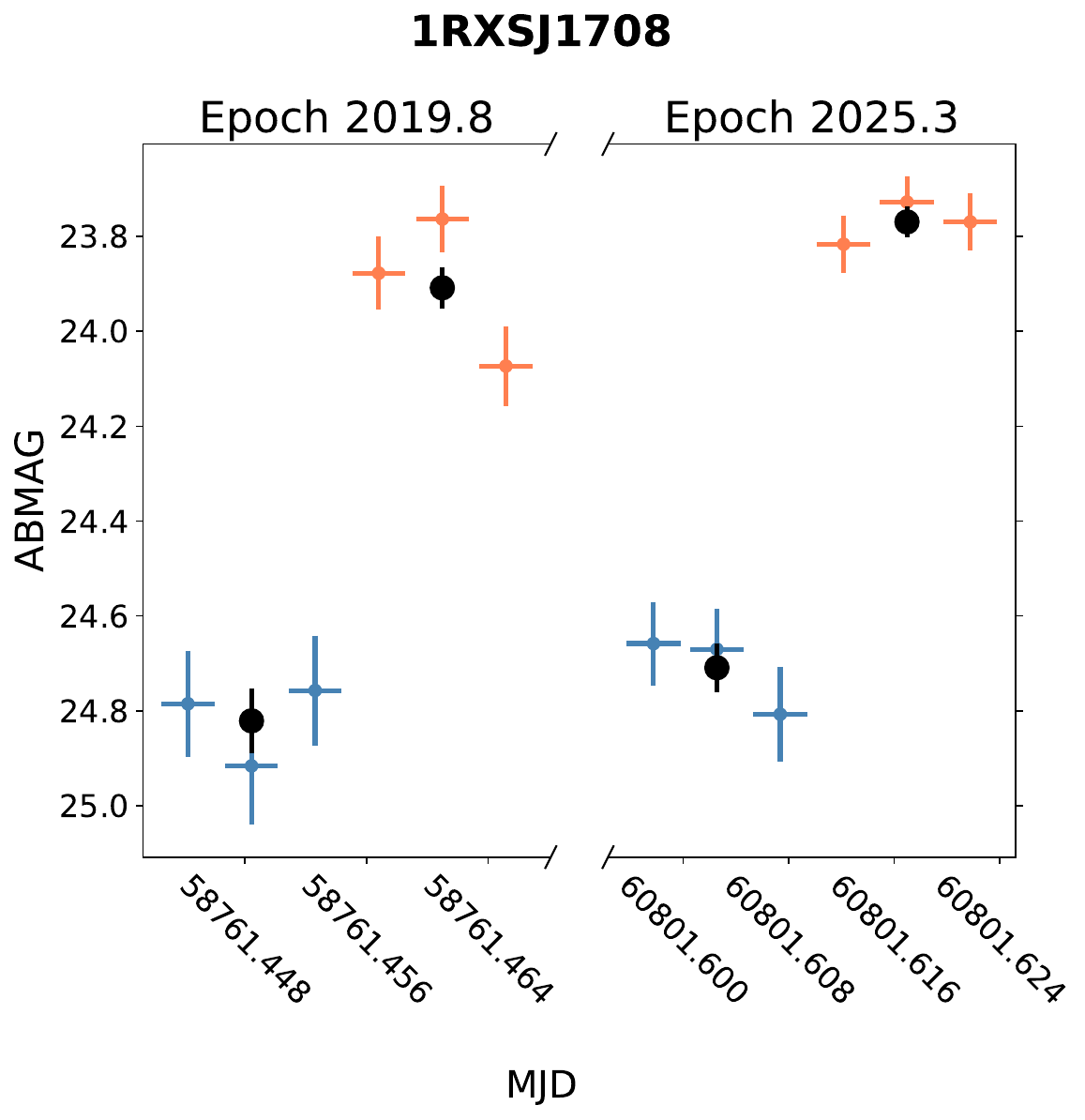}
\includegraphics[width=0.9\textwidth]{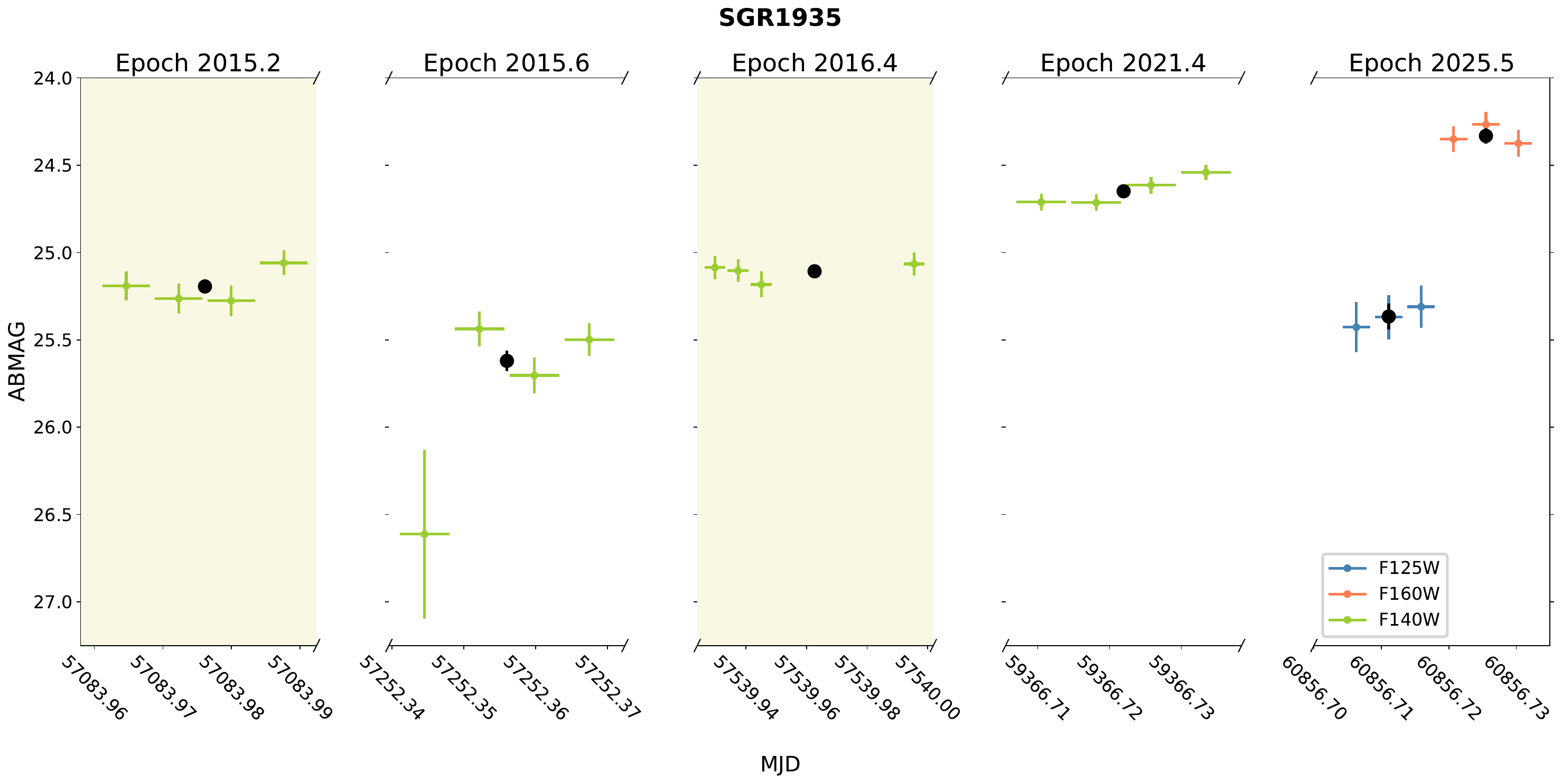}
\caption{Near-IR HST {\sc \_flt}-level light-curves for the six counterparts identified in our sample. F125W apparent magnitudes are coloured blue, F160W (which are exclusively fainter than F125W) orange and F140W green (the first four epochs for SGR1935 only). Note that each subplot has different y-axis limits. Values are reported in the AB system and are not corrected for Galactic extinction. In each epoch, photometry on the combined, drizzled  {\sc \_drz} image is also show as a black point with error-bars at the midpoint of the {\sc \_flt} observations. The 2015.2 and 2016.4 panels for SGR1935 are shaded to indicate that the magnetar was in outburst during these observations (see Section \ref{sec:outbursts}). }
\label{fig:LC}
\end{figure*}

\begin{figure*}
\centering
\includegraphics[width=0.9\textwidth]{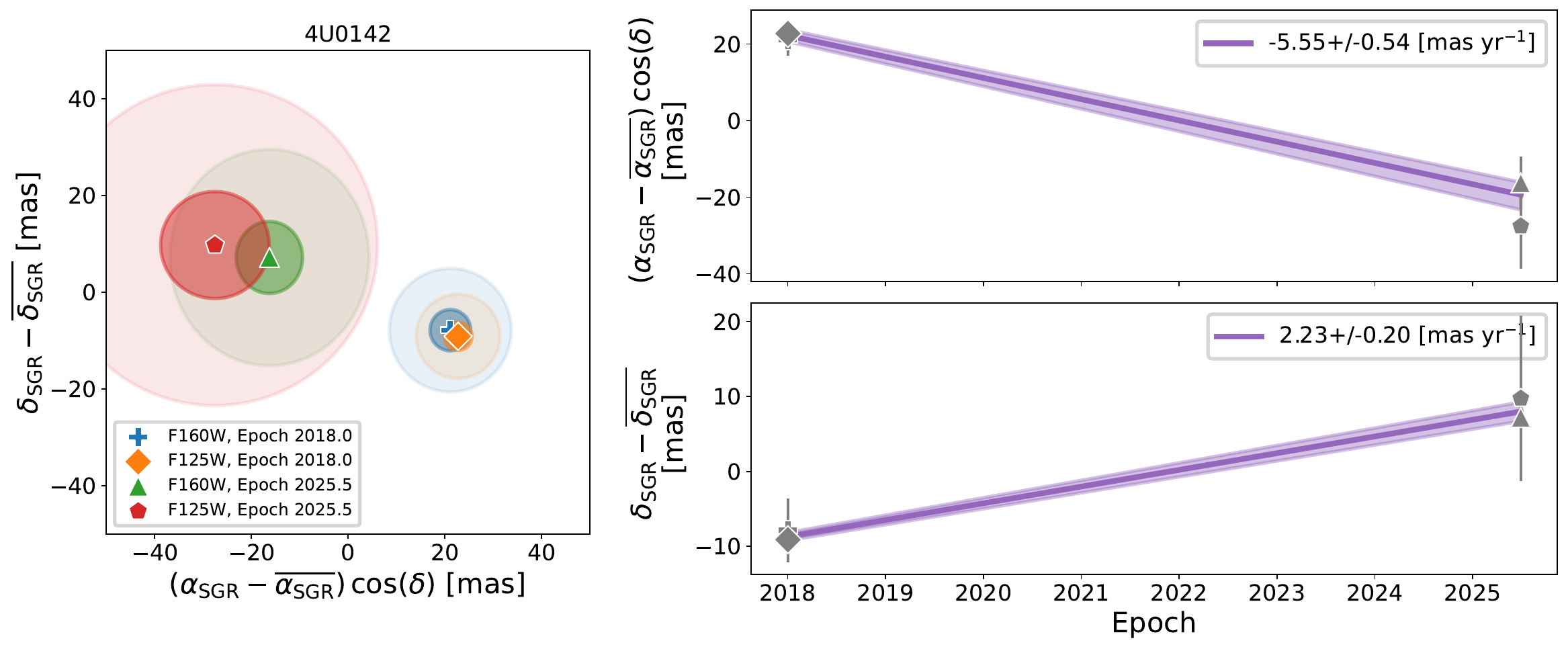}
\includegraphics[width=0.9\textwidth]{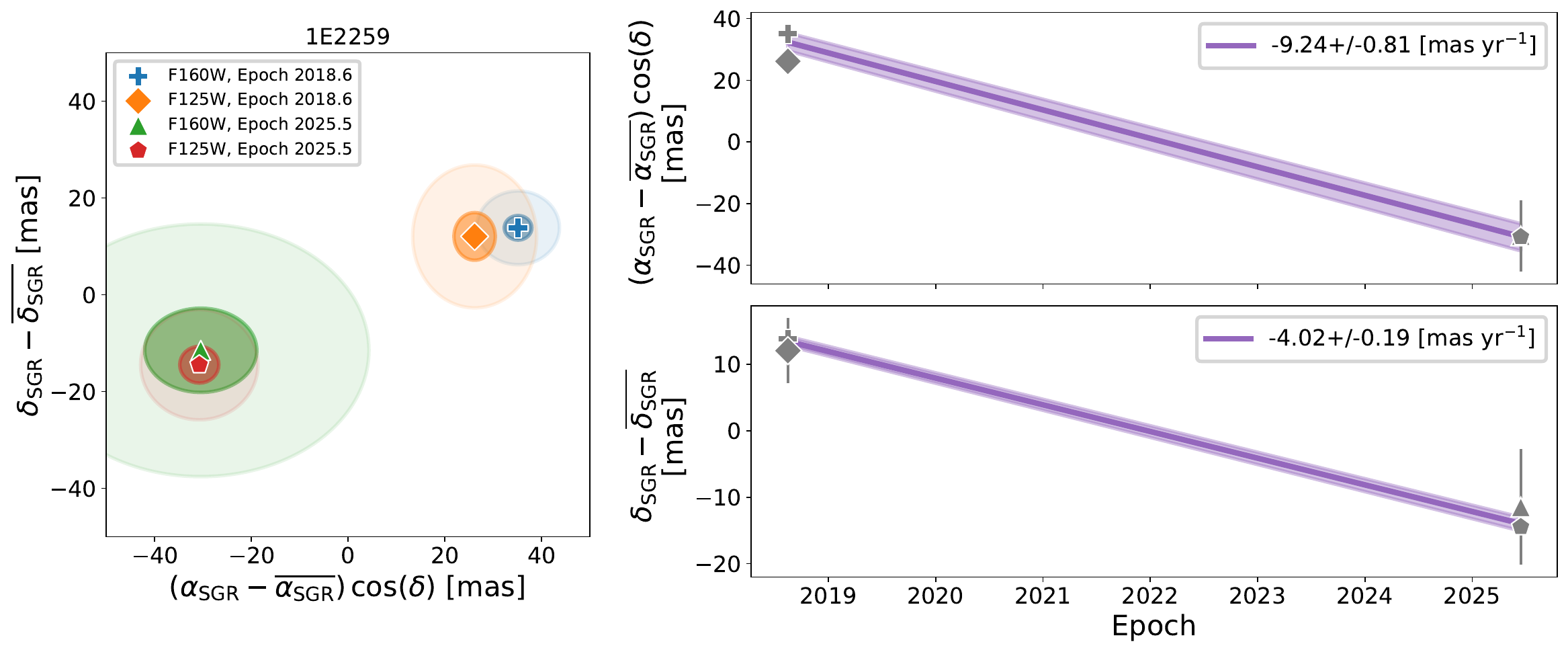}
\includegraphics[width=0.9\textwidth]{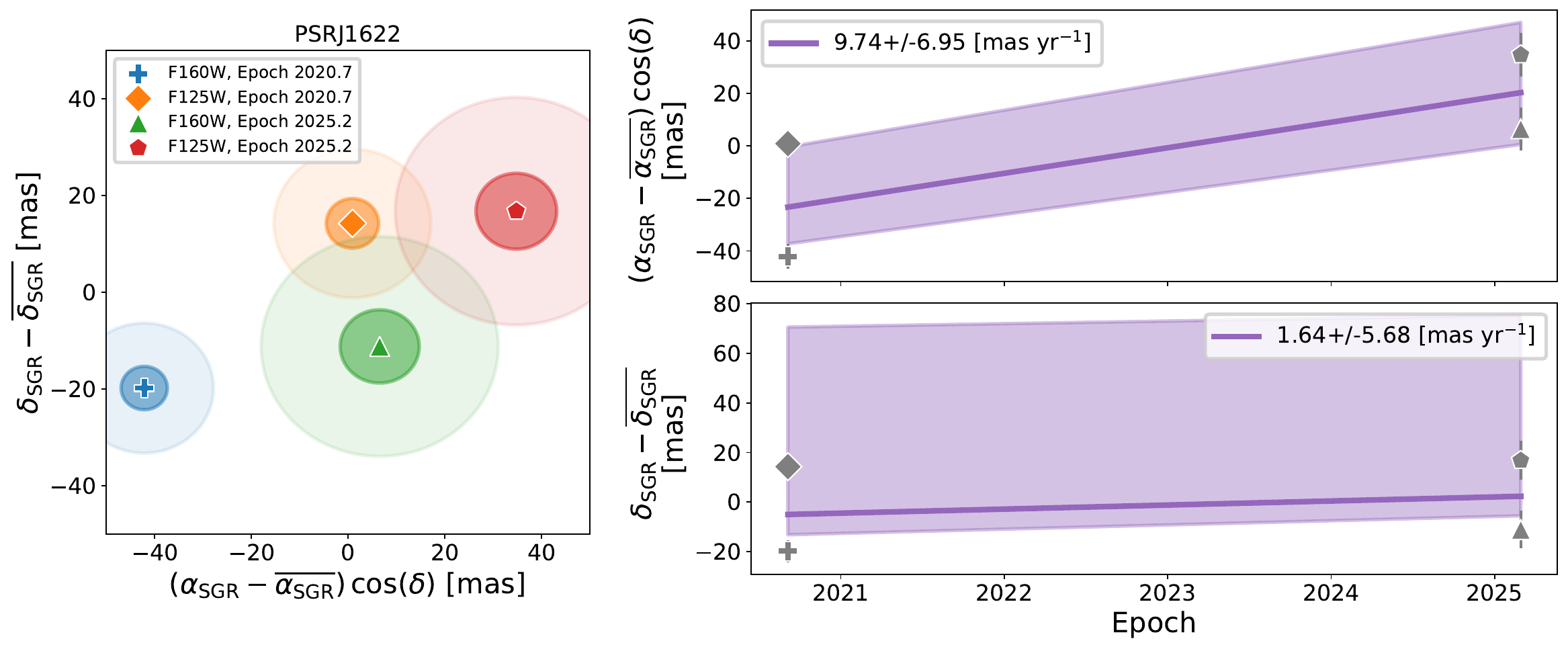}
\caption{Proper motion measurements for 4U0142, 1E2259 and PSRJ\,1622. Left panels: IR counterpart offsets in $\alpha$,$\delta$ with respect to the mean position across the four epochs, and the 1\,$\sigma$ (dark shading) and 3\,$\sigma$ (lighter shading) uncertainties on these offsets. Right panels: peculiar proper motions determined from each epoch and the best fit to all four measurements. The symbols match the epochs/filters shown in the left-hand panel. The proper motion in declination is poorly constrained for PSRJ1622, and could also be expressed as a upper limit on the proper motion of $\mu_{\delta} < 11.2$\,mas\,yr$^{-1}$ (at 2$\sigma$ confidence). }
\label{fig:pmfit1}
\end{figure*}
\afterpage{\clearpage}

\begin{figure*}
\centering
\includegraphics[width=0.9\textwidth]{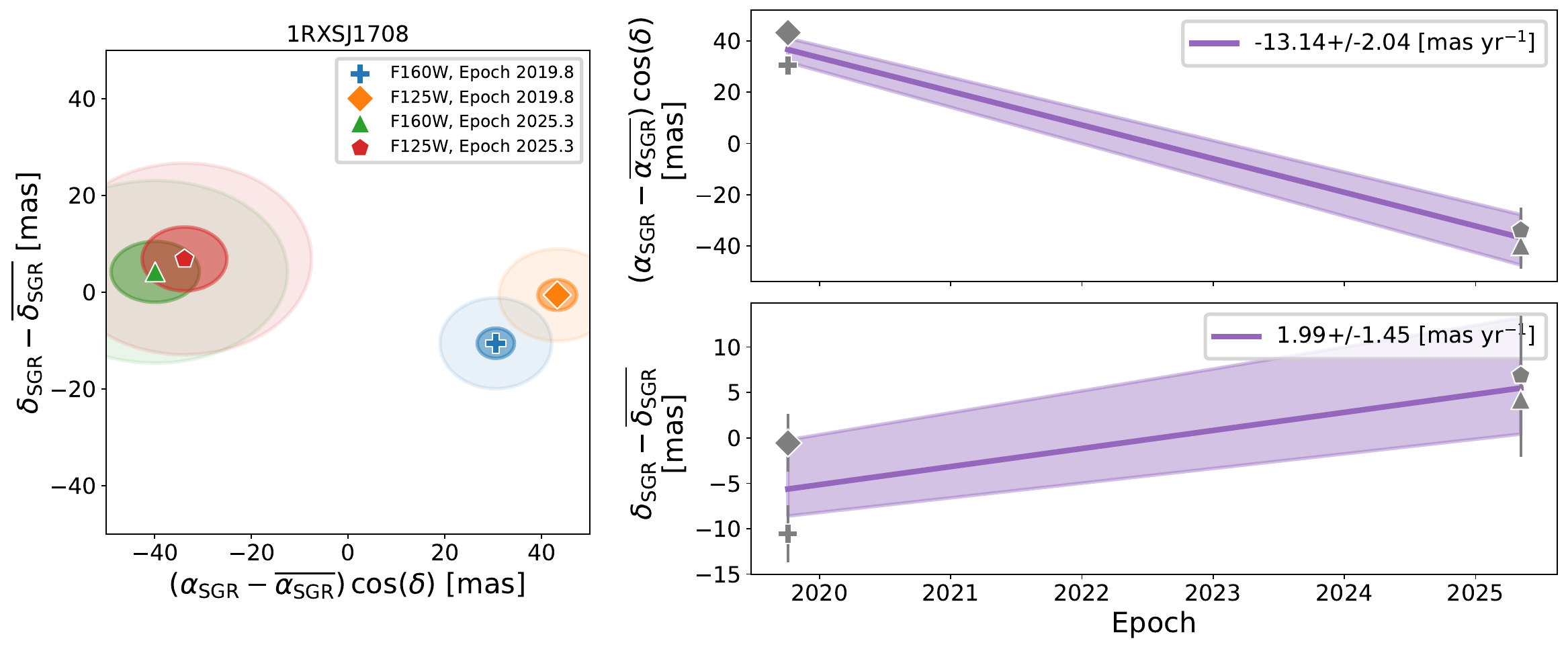}
\includegraphics[width=0.9\textwidth]{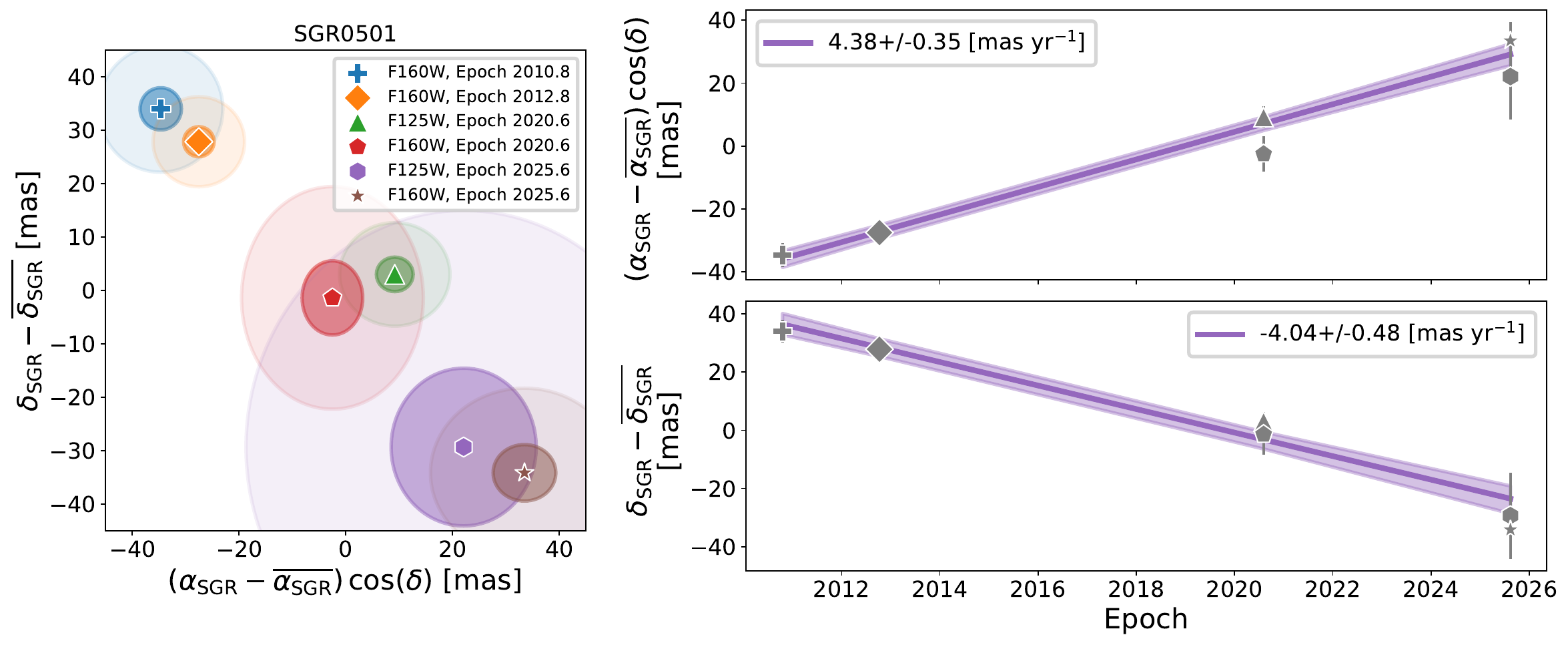}
\includegraphics[width=0.9\textwidth]{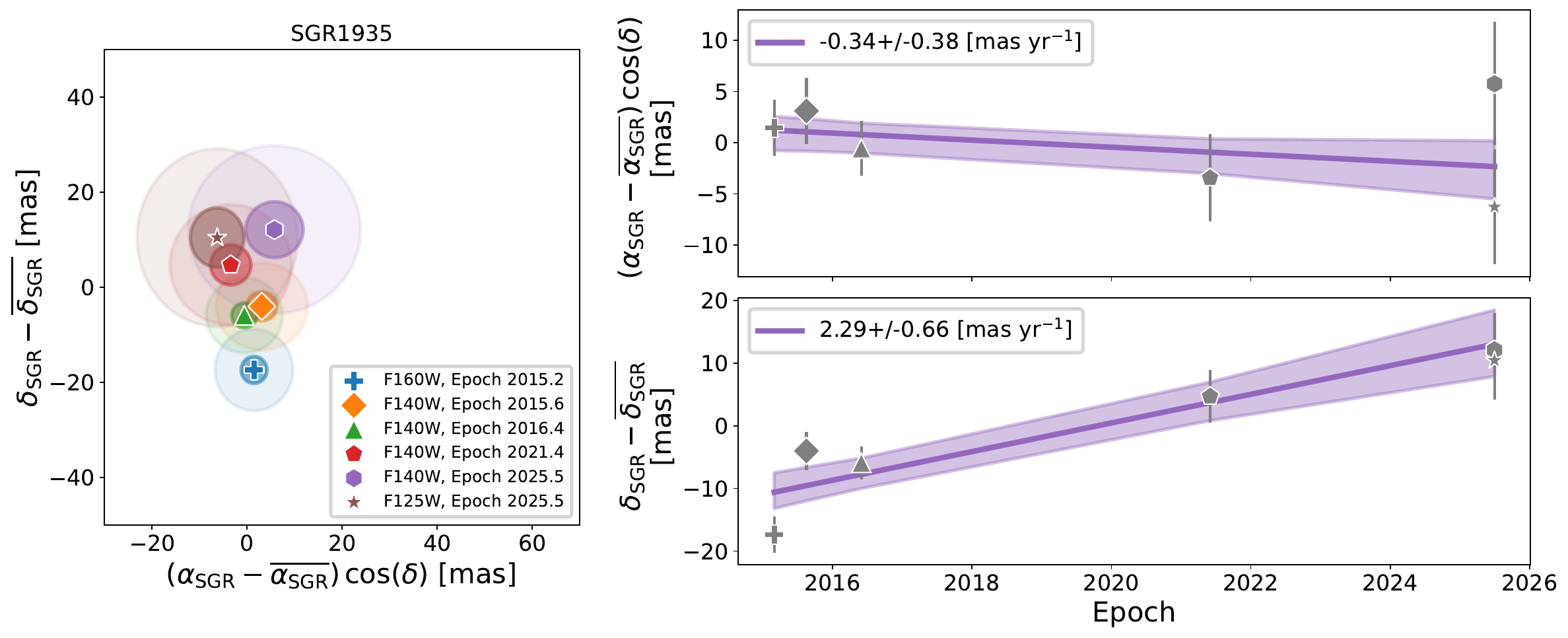}
\caption{Continued as in Figure \ref{fig:pmfit1} for magnetars 1RXSJ\,1708, SGR0501 and SGR1935. }
\label{fig:pmfit2}
\end{figure*}
\afterpage{\clearpage}

\subsubsection{1RXS\,J170849-400910}\label{sec:1708}
We identify a new candidate counterpart to 1RXS\,J1708 based on its unusual kinematics. This is not the first counterpart candidate for 1RXS\,J1708, and it is neither of the brighter sources near the south-eastern edge of the X-ray error circle identified by \citet{2003ApJ...589L..93I,2006ApJ...648..534D}. Although the counterpart claimed by \citet{2006ApJ...648..534D} showed signs of variability (at 3$\sigma$ significance), \citet{2022MNRAS.513.3550C} showed that this source would have an intrinsic $H$-band absolute magnitude of $\sim$5, far brighter than that of securely identified counterparts (e.g. 4U0142, SGR0501). The 3.8\,kpc distance to this magnetar is based on the optical extinction, which was first estimated from absorption edges in the magnetar X-ray spectra \citep[and is consistent with the A${\rm _V}$-N$_{\rm H}$ relation][]{2022MNRAS.513.3550C}, and the distance through the Galactic plane that this extinction implies \citep{2006ApJ...648..534D,2006ApJ...650.1070D}. At this distance, and with extinction A$_{\rm V}=7.6$, the new counterpart identified in this work would have absolute magnitude of $\sim$8. This is closer to the known population, and coupled with its discrepant kinematics with respect to field, we deem the faint source labelled in Figure \ref{fig:stamps} to be the likely counterpart. The IR efficiency of this source would then then be log$_{10}$(L$_{\rm IR}$/$\dot{E}$) = -1.3, adopting $\dot{E} = 0.58\times 10^{33}$\,erg\,s$^{-1}$ \citep{2014ApJS..212....6O}.

\subsubsection{CXOUJ164710.2-455216}
\begin{figure*}
\centering
\includegraphics[width=0.99\textwidth]{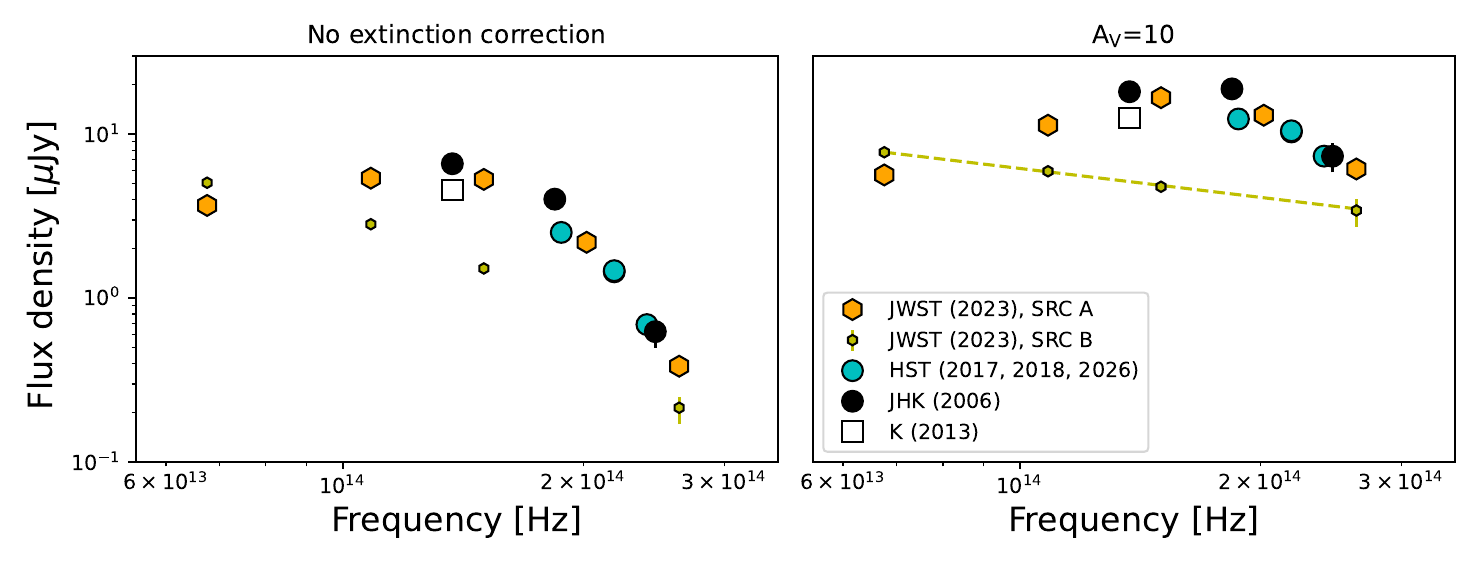}
\caption{The NIR spectral energy distribution (SED) of the two objects close to the centre of the X-ray localisation of CXOUJ1647. One is a source previously identified as a candidate counterpart \citep{2008A&A...482..607T}, however the other has a power-law SED after extinction correction, similar to other known magnetar counterparts, so we deem this the likely counterpart. Measurements of the brighter, previously identified source are from \citet[][for the 2006 JHK$_{\rm s}$ and 2013 K$_{\rm s}$ points]{2018MNRAS.473.3180T}, this work and \citet[][HST]{2022MNRAS.512.6093C} and \citet[][for the JWST/NIRcam data]{2025A&A...693A.120G}.  Note that in Figure \ref{fig:cp_ident_no2}, neither of the objects in the X-ray error circle with a proper motion measurement are the power-law source, since it is not detected in the HST images.}
\label{fig:CXOUJ1647}
\end{figure*}

Finally, we report the likely identification of the infrared counterpart of magnetar CXOUJ1647, which lies outside the Westerlund 1 massive star cluster \citep{2006ApJ...636L..41M}. None of the HST sources in the X-ray error circle pass our counterpart selection by the stated criteria. This includes the object identified as a candidate counterpart by \citet{2018MNRAS.473.3180T}, which we will refer to as source A. However, in the deeper JWST imaging available for the field, a second source (B) is detected close to the centre of the error circle. We find that the extinction-corrected ($A_{\rm V}=10$) SED of source B is well described by a power-law of the form f$_{\nu} \propto \nu^{-\alpha}$, with $\alpha = -0.5854\pm0.0003$. This is in contrast with $\alpha=2$ for the Rayleigh-Jeans tail of a black body expected for stellar SEDs. Given that power-law SEDs appear to be typical for magnetars at these wavelengths \citep[][]{2016MNRAS.458L.114M,2024ApJ...972..176H}, the otherwise typically stellar SED of source A, and the unremarkable colours of the other candidates in the X-ray error circle, we deem source B the likely counterpart and label it as such in Figure \ref{fig:stamps}. The SEDs of sources A and B are shown in Figure \ref{fig:CXOUJ1647}, but we leave a full multi-wavelength study of this magnetar to future work (Borghese et al., in prep). For the purposes of this paper, as source B is only detected in the JWST imaging, no proper motion measurement is possible without a second JWST epoch. This is one of only a few of magnetars thus far with a candidate former binary companion \citep{2014A&A...565A..90C,2024MNRAS.531.2379S}, and the proper motion of this candidate former companion is well constrained by {\em Gaia}, so determining the past trajectory of CXOUJ1647 is an important and viable test of their association.

\subsection{Other non-detections}\label{sec:nodetect}
There are 8 magnetars in our sample of 14 for which we fail to identify a good counterpart candidate in the HST imaging. Of these, SGR1900 and CXOU1647 have candidates which are detected in previous HST/JWST imaging as discussed above, leaving six with no known counterpart. In these cases, non-detections can occur for two reasons: either the counterpart is not detected in the images, or it is detected, but does not exceed our 3$\sigma$ selection criteria with respect to other sources in the field in terms of colour, proper motion or variability. We list upper limits on the flux, colour and variability for the six magnetars with no counterpart candidate 
in Table \ref{tab:mags}. Similarly, upper limits on the proper motion are provided in Table \ref{tab:mu}. Note these proper motion upper limits do not directly reflect the minimum measurable proper motion, rather they are minimum proper motions that would be required for an object to fall outside the 3$\sigma$ range of measured proper motions in the image.

Sources in the fields without detections are typically redder than in the fields with identified counterparts, as we can see from the upper limits on the colour for the non-identified sources in Table \ref{tab:mags}. Along these dustier sight-lines, there are more red sources and so identification through a discrepant colour is less effective. Likewise, upper limits on velocity in the non-detected cases are much higher than the typical values of the known counterparts (see Table \ref{tab:mu} and Figure \ref{fig:vt}). It is therefore plausible that counterparts to SWIFTJ1822, SWIFTJ1834, SGR1627, CXOUJ1714, SGR1833 and AXJ1818 are detected in the HST imaging, but without exhibiting high variability, excessively red colours or high proper motions. In this scenario we have no way to confidently identify them. On the other hand, higher distances and extinctions which lead to high upper limits on colour and velocity will also tend to make these counterparts harder to detect, making it hard to distinguish between non-detections, versus non-identifications.

\subsection{Light curves and correlation with high energy activity}\label{sec:outbursts}
Near-IR HST light-curves for the six counterpart candidates detected in the HST imaging are provided in Figure \ref{fig:LC}, including photometry on the individual {\sc \_flt} level images, which have undergone calibrations including flat fielding and dark subtraction, but have not yet been combined with the drizzling process. Some level of variability is apparent in most of these sources, even those which were not selected on this criteria (e.g. SGR0501). Another example is 4U0142, which displays spectral variability, apparently transitioning from a power-law plus blackbody to a pure power-law in the optical-IR regime over the course of 20 years \citep{2006Natur.440..772W,2024ApJ...972..176H}. However, in the near-IR, the flux only varied at the $\sim$10-20\% level, again demonstrating that IR spectral energy distributions (only possible with JWST) reveal much that observations in only 1-2 filters miss. 

We searched the literature for reports of flares and outbursts near the epochs of observation shown in Figure \ref{fig:LC}. 1RXS\,J1708 produced a flare on 28 August 2018 \citep{2020ApJ...889L..27Y}, around 2 months before the 2018 HST epoch, but has otherwise not gone into outburst in recent years. CXOUJ1714 produced a burst on 5 February 2018, several months before the 2018 HST epoch, but again has otherwise not gone into outburst since. 

SGR1935 has been active in recent years. The first 2015 HST epoch for SGR1935 occurred during an outburst which began on 22 February 2015, while the 2016 HST imaging was taken shortly after the onset of an outburst on 18 May 2016 \citep{2017ApJ...847...85Y}, see also \citet{2018ApJ...854..161L} and \citet{2022ApJ...926..121L}. The 2021 epoch of imaging was taken in between two bursts in February and July of that year \citep[e.g.][]{2021ATel14388....1B,2021GCN.30407....1H}, we but note that the source went into outburst in April 2020 and October 2022 \citep[also associated with fast radio burst emission,][]{2020Natur.587...59B,2020Natur.587...54C,2021NatAs...5..401T,2021NatAs...5..414K,2024ApJ...965...87I}, with frequent bursts in the period in-between. During the final 2025 epoch, the source was in quiescence.  We indicate the epochs where the source was active in Figure \ref{fig:LC} (i.e. 2015.2 and 2016.4) by shading the background of the relevant panels. There is no obvious correlation between X-ray activity and NIR flux. For a more detailed analysis of the correlation between SGR1935's X-ray and NIR variability, we refer the reader to \citet{2022ApJ...926..121L}, who also note that the NIR variability does not seem to track activity at high energies. 

Finally, CXOUJ1647 was active from May 2017 through to at least April 2018, producing multiple bursts in this time \citep{2019MNRAS.484.2931B}, but as discussed above, the likely NIR counterpart was not detected in any of the HST epochs. Analysis of the X-ray light curve of this source from 2023 onwards in ongoing. No bursts, outbursts or flares were found near any of the epochs for the other magnetars in the sample.

\begin{table*}
	\centering
	\caption{Photometry of the NIR counterparts reported in this work (not corrected for Galactic extinction). For details of the observations, and previous HST epochs in these filters, see Table \ref{tab:obsv} and Figure \ref{fig:LC}. For SGR0501 we use the 2020.6 epoch for the F160W $\Delta$mag. The counterpart of CXOUJ1647 is not detected in the HST imaging, we list here the JWST/NIRcam photometry of \citet{2025A&A...693A.120G}. Below the horizontal line, we list 3$\sigma$ limiting magnitudes and fluxes for the six objects in this sample where there is no good counterpart candidate known (i.e. the eight shown in Figures \ref{fig:cp_ident_no} and \ref{fig:cp_ident_no2}, minus CXOUJ1647 which was detected with JWST, and SGR1900, see Section \ref{sec:1900}). We also list 3$\sigma$ upper limits on the F125W-F160W colour and $\Delta$mag, which a source would have to exceed in order to be selected as a candidate following the criteria outlined in Section \ref{sec:ident}. We are unable to say whether a counterpart was undetected, versus detected but not selected, in these cases. }
	\label{tab:mags}
	\begin{tabular}{lcccccc} 
		\hline
		  \hline
            Magnetar & Epoch & Filter & m(AB) & Flux [$\mu$Jy] & F125W-F160W & |$\Delta$mag| \\
            \hline						  
            4U0142	&	2025.5	&	F125W	&	22.71$\pm0.01$	&	2.99$\pm$0.03	&	0.54$\pm$0.01	&	0.04$\pm$0.02		\\
            	&	2025.5	&	F160W	&	22.17$\pm0.01$	&	4.92$\pm$0.05	&		&	0.07$\pm$0.02		\\
            SGR0501	&	2025.6	&	F125W	&	24.64$\pm0.04$	&	0.51$\pm$0.02	&	0.52$\pm$0.06	&	0.39$\pm$0.06		\\
            	&	2025.6	&	F160W	&	24.12$\pm0.04$	&	0.82$\pm$0.03	&		&	0.38$\pm$0.05		\\
            PSRJ1622	&	2025.2	&	F125W	&	24.47$\pm0.05$	&	0.59$\pm$0.03	&	0.75$\pm$0.06	&	0.10$\pm$0.06		\\
            	&	2025.2	&	F160W	&	23.52$\pm0.03$	&	1.42$\pm$0.04	&		&	0.47$\pm$0.05		\\
            SGR1935	&	2025.5	&	F125W	&	25.37$\pm0.08$	&	0.26$\pm$0.02	&	1.04$\pm$0.09	&			\\
            	&	2025.5	&	F160W	&	24.33$\pm0.04$	&	0.67$\pm$0.02	&		&			\\
            1E2259	&	2025.4	&	F125W	&	23.60$\pm0.02$	&	1.32$\pm$0.02	&	0.48$\pm$0.03	&	 1.09$\pm$0.06		\\
            	&	2025.4	&	F160W	&	23.12$\pm0.02$	&	2.05$\pm$0.04	&		&	 0.81$\pm$0.05		\\
            1RXSJ\,1708	&	2025.3	&	F125W	&	24.71$\pm0.05$	&	0.47$\pm$0.02	&	0.94$\pm$0.06	&	 0.11$\pm$0.09		\\
            	&	2025.3	&	F160W	&	23.77$\pm0.03$	&	1.13$\pm$0.03	&		&	 0.14$\pm$0.06		\\													
            CXOUJ1647	&	2023.2	&	F115W	&	25.57$\pm$0.20	&	0.21$\pm$0.04	&		&			\\
            	&	2023.2	&	F200W	&	23.45$\pm$0.04	&	1.51$\pm$0.05	&		&			\\
            	&	2023.2	&	F277W	&	22.77$\pm$0.02	&	2.83$\pm$0.05	&		&			\\
            	&	2023.2	&	F444W	&	22.14$\pm$0.01	&	5.06$\pm$0.05	&		&			\\
            \hline	
            SWIFTJ1822	&	2025.2	&	F125W	&	$<$25.6	&	$<$0.21	&	$<$1.30	&	$<$0.27		\\
            	&	2025.2	&	F160W	&	$<$25.4	&	$<$0.24	&		&	$<$0.26		\\
            SWIFTJ1834	&	2025.2	&	F125W	&	$<$26.7	&	$<$0.07	&	$<$3.31	&	$<$0.22		\\
            	&	2025.2	&	F160W	&	$<$26.3	&	$<$0.11	&		&	$<$0.23	\\
            SGR1627 &	2025.3	&	F125W	&	$<$26.4	&	$<$0.10	&	$<$2.40	&	$<$0.26		\\
            	&	2025.3	&	F160W	&	$<$25.9	&	$<$0.16	&		&	$<$0.24	\\
            CXOUJ1714 &	2025.3	&	F125W	&	$<$25.2	&	$<$0.31	&	$<$1.70 &	$<$0.26		\\
            	&	2025.3	&	F160W	&	$<$24.4	&	$<$0.65 &		&	$<$0.25	\\
            SGR1833 &	2025.5	&	F125W	&	$<$26.4	&	$<$0.10	&	$<$3.49	&	$<$0.26		\\
            	&	2025.5	&	F160W	&	$<$25.9	&	$<$0.16	&		&	$<$0.26	\\
            AXJ1818 &	2025.4	&	F125W	&	$<$26.4	&	$<$0.10	&	$<$1.88	&	$<$0.29		\\
            	&	2025.4	&	F160W	&	$<$25.9	&	$<$0.16	&		&	$<$0.27	\\
        \hline	
	\end{tabular}
\end{table*}

\subsection{Transverse velocities}\label{sec:vt}
For the six NIR counterpart identifications with proper motion measurements, figures visualising the fits are given in Figures \ref{fig:pmfit1} and \ref{fig:pmfit2}. In each of the four cases where we measure the proper motion of an object which already has a previous measurement, our results are fully consistent with the previous results (listed in Table \ref{tab:mu}), with uncertainties typically reduced by a factor of $\sim$2. We convert peculiar proper motions into transverse velocities by adopting the distances listed in Table \ref{tab:mu}. Our six magnetar velocities are supplemented by XTE\,J1810 \citep{2020MNRAS.498.3736D}, 1E1547 \citep{2012ApJ...748L...1D}, SGR1806 and SGR1900 \citep{2012ApJ...761...76T} and SWIFTJ1818 \citep{2024ApJ...971L..13D}, for a total of 11 magnetars with transverse velocity measurements. The proper motions, distances and velocities of these additional six magnetars are provided in Table \ref{tab:mu}. The transverse velocity of SGRJ1745 has also been measured \citep{2015ApJ...798..120B}, but it is likely bound to Sgr A$^{\star}$, so we do not consider it further. 

The cumulative distribution of the 11 magnetar velocities is shown in Figure \ref{fig:vt}. We generate 1000 realisations of the distribution by sampling the proper motion uncertainties, assuming they are Gaussian. We note that distance uncertainties are not accounted for, and although these can be considerable \citep[only two magnetars have a parallax measurement,][]{2020MNRAS.498.3736D,2024ApJ...971L..13D}, they are hard to formally quantify. Each step in the distribution is labelled by the method through which the best available proper motion was measured. We note that parallaxes are unlikely to strongly influence our proper motion measurements: for even the closest magnetars ($\sim$1\,kpc), the maximum positional shift due to parallax would be $\sim$1\,mas, and hence $<$0.2\,mas\,yr$^{-1}$ over the baselines probed, if such a shift were mis-attributed to proper motion. This is typically a factor of $\sim$10 or more lower than the proper motions measured (see Table \ref{tab:mu}).

\begin{table*}
	\centering
	\caption{Proper motion, distance and transverse velocity information for the sample of 11 magnetars with these measurements as defined in Section \ref{sec:vt}. The table is split into four sections. Top section: peculiar proper motions for the six NIR counterparts (re-)identified in this work. The proper motions of PSRJ1622 and 1RXS\,J1708 are newly reported. Uncertainties on the v$_{t}$ velocities do not include uncertainties on the distance, while the final column lists velocities v$_{\rm t,d}$ which also include the quoted distance uncertainties (a 20\% distance uncertainty is adopted for PSRJ\,1622). Second section: previous measurements for the four previously known counterparts we re-identified in this work. Third section: 3$\sigma$ upper limits on proper motion and velocity in the six cases where no counterpart has been identified, under the assumption that the counterpart was detected in the image, but not discrepant in terms of colour, variability or proper motion. Final section: previously reported proper motion measurements, distance estimates and transverse velocities for five additional magnetars which have such measurements in the literature. Velocity uncertainties include the quoted distance uncertainty in each case.  }
	\label{tab:mu}
	\begin{tabular}{lllllll} %
		\hline
		  \hline
            Magnetar	& $\mu_{\alpha}$cos($\delta$)	&  $\mu_{\delta}$ & $d$  & Distance ref.	&	$v_{\rm t}$ & $v_{\rm t,d}$	\\
            	& [mas\,yr$^{-1}$]	&   [mas\,yr$^{-1}$] & [kpc]  &    &    [km/s]	&	[km/s]	\\
            \hline										
            PSRJ1622	&	9.7$\pm$7.1	&	1.6$\pm$5.8	&		$\sim$5	&	\citet{2017ApJ...835...29Y}$^\star$	&	260$\pm$180	& 260$^{+198}_{-192}$ \\
            1E2259	&	-9.2$\pm$0.8	&	-4.0$\pm$0.2	&	3.2$\pm$0.2	&	\citet{2012ApJ...746L...4K}	&	153$\pm$11	& 153$\pm$14 \\
            4U0142	&	-5.5$\pm$0.5	&	2.2$\pm$0.2	&		3.6$\pm$0.4	&	\citet{2006ApJ...650.1070D}	&	102$\pm$9	& 102$\pm$15 \\
            SGR0501	&	4.4$\pm$0.4	&	-4.0$\pm$0.5	&		2$\pm$1	&	\citet{2011ApJ...739...87L}	&	57$\pm$4	& 57$\pm$31\\
            SGR1935	&	-0.3$\pm$0.4	&	2.3$\pm$0.7	&		6.6$\pm$0.7$\dagger$	&	 \citet{2020ApJ...905...99Z}	&	72$\pm$20	& 72$\pm$21\\
            1RXSJ\,1708	&	-13$\pm$2	&	2.0$\pm$1.5	&	3.8$\pm$0.5	&	 \citet{2006ApJ...650.1070D}	&	240$\pm$40	& 240$\pm$50 \\
            \hline	
            Magnetar	 & $\mu_{\alpha, {\rm prev}}$cos($\delta$)	&	$\mu_{\delta, {\rm prev}}$ & &  $\mu_{\rm prev}$ ref.		\\
               & [mas\,yr$^{-1}$]	&   [mas\,yr$^{-1}$] &   &    	\\
            \hline	
            1E2259		&	-9.9$\pm$1.1	&	-3.0$\pm$1.1	&	& \citet{2013ApJ...772...31T}		&		\\
            4U0142	&	-5.6$\pm$1.3	&	2.9$\pm$1.3	&	& \citet{2013ApJ...772...31T}		&		\\
            SGR0501	&	4.1$\pm$0.7	&	-3.5$\pm$0.4	&	& \citet{2025AandA...696A.127C}		&	\\
            SGR1935	&	-0.7$\pm$0.8	&	 3.1$\pm$1.6	&	& \citet{2022ApJ...926..121L}$\dagger$		&		\\
           \hline	
            Magnetar	& $\mu_{\alpha}$cos($\delta$)	&  $\mu_{\delta}$ & $d$  & Distance ref.	&	$v_{\rm t}$ & 	\\
            	& [mas\,yr$^{-1}$]	&   [mas\,yr$^{-1}$] & [kpc]  &    &    [km/s]	&	\\
            \hline		
            SWIFTJ1822 &	$<$5.7	&	$<$7.5	&		1.6$\pm$0.3	& \citet{2012ApJ...761...66S}	       &	$<$71	&  \\
            SWIFTJ1834 &	$<$6.1	&	$<$7.5	&		4.2$\pm$0.3	&	\citet{2008AJ....135..167L}       &	$<$193	&  \\
            SGR1627 &	$<$5.8	&	$<$5.9	&		11.0$\pm$0.3	&	\citet{1999ApJ...526L..29C}       &	$<$432	&  \\
            CXOUJ1714 &	$<$5.5	&	$<$6.0	&		9.8$\pm$1.5	&	\citet{2019MNRAS.487.5019B}     &	$<$380	&  \\
            SGR1833 &	$<$5.1	&	$<$6.9	&		$<10^{\star \star}$	&	       &	$<$408	&  \\
            AXJ1818 &	$<$6.4	&	$<$8.9	&		$<10^{\star \star}$	       &    &	$<$521	&  \\
            \hline 
           Magnetar  & $\mu_{\alpha, {\rm prev}}$cos($\delta$)	&	$\mu_{\delta, {\rm prev}}$  	& $d$  &   $\mu_{\rm prev}$, $d$ and $v_{\rm t,d}$ ref.	 & & $v_{\rm t, d}$ 	  \\
               & [mas\,yr$^{-1}$]	&   [mas\,yr$^{-1}$] &  [kpc]  &    & & [km/s]	\\
            \hline	
                1E1547		&	4.8$\pm$0.5	&	 -7.9$\pm$0.3	&	6$\pm$2	&	\citet{2012ApJ...748L...1D} & & 280$^{+130}_{-120}$		\\
                SGR1806		&	4.2$\pm$0.9	&	 7.0$\pm$1.8	&	9$\pm$2	&	\citet{2012ApJ...761...76T} & & 350$\pm$100 		\\
                SGR1900		&	2.7$\pm$0.2	&	4.8$\pm$0.4	&	12.5$\pm$1.6	& \citet{2012ApJ...761...76T} & &	130$\pm$30	\\
                XTE\,J1810		&	-3.79$^{+0.05}_{-0.03}$	&	 -16.2$\pm$0.1	& 	 2.5$^{+0.4}_{-0.3}$	& \citet{2020MNRAS.498.3736D} &	 &	198$^{+29}_{-23}$		\\
                SWIFTJ1818		&	-3.57$\pm$0.02	&	-7.72$\pm0.06$	&	9.4$^{+2.0}_{-1.4}$	&	\citet{2024ApJ...971L..13D}	 & & 48$^{+50}_{-16}$ \\

            \hline
	\end{tabular}
    \newline
    $^\star$ - See also Section \ref{sec:1622}. $\dagger$ - See also \citet{2021MNRAS.503.5367B}. $^{\star \star}$ - No distances are available, we assume $d<10$\,kpc. 
\end{table*}

\subsection{Birth sites}\label{sec:bsite}
For the six counterparts with proper motions identified in Section \ref{sec:cps}, we produce 100 realisations of their past trajectories by sampling from the proper motion uncertainties and plotting these vectors with lengths corresponding to (i) their characteristic age \citep[from the McGill catalogue][, and references therein]{2014ApJS..212....6O} and (ii) an age ten times older than this to account for the possibility of inaccurate characteristic ages. We note that characteristic ages are typically upper limits, but that they could be higher if $\dot{P}$ was measured outside quiescence, when spin-down rates can vary significantly \citep[e.g.][]{2002ApJ...576..381W,2003ApJ...596L..71K,2013ApJ...770L..23M}. Since the influence of the Galactic potential is not expected to significantly affect the trajectory of natal neutron stars on timescales of less than $\sim$10\,Myr \citep{2024A&A...687A.272D}, the linear propagation of past trajectories is expected to be reliable for our purposes. The results are shown in Figure \ref{fig:traject}.

SGR1935 can be traced back, within a factor of a few of its characteristic age, to the geometric centre of SNR G57.2+08 as first demonstrated by \citet{2022ApJ...926..121L}. 1E2259 also has a past trajectory which passes close to the centre of a SNR \citep[CTB 109,][]{1981Natur.293..202F,2013ApJ...772...31T}, but the magnetar must be significantly younger than its characteristic age which otherwise places the birth site well outside the SNR. As first shown by \citet{2025AandA...696A.127C}, the direction and magnitude of SGR0501's proper motion rules out an origin in the nearby SNR HB9, and no other plausible birth site has yet been identified. Similarly, no candidate SNR or stellar cluster has been found along the past trajectory of 4U0142 \citep[see also][]{2013ApJ...772...31T}. SNR G127.1$+$0.5 lies within 2\,deg of 4U0142, but in the opposite direction. The past trajectory of PSRJ1622 is consistent, within a factor of $\sim$2 of its characteristic age, with the centre of SNR\,G333.9$+$0.0 \citep[first noted as a possible origin of this magnetar by][]{2012ApJ...751...53A}.

Finally, the past trajectory of the object we claim to be the counterpart of 1RXS\,J1708 intersects SNR G346.6-0.2. There is no other neutron star association with this SNR \citep{2025JApA...46...14G}. 1RXS\,J1708 would have to be $\sim$10 times its characteristic age in order for this association to hold, although such a large kinematic age is not unprecedented - XTEJ1810 would have to be $\sim$70\,kyr old to be associated with SNR\,G11.0-0.0 \citep{2007ApJ...662.1198H,2020MNRAS.498.3736D}, and similar discrepancies have been seen in pulsars \citep[e.g.][]{2022A&A...659A..41E}. Despite the past spatial coincidences, the reliability of these associations is called into question by the large discrepancies between the SNR and kinematic ages. Furthermore, distance estimates for SNR G346.6-0.2 are in the range $\sim$5.5-11\,kpc \citep{1998AJ....116.1323K,2021ApJ...923...15S}, higher than the estimate of 3.8$\pm$0.5 claimed for 1RXS\,J1708 based on red clump stars along that sight-line \citep{2006ApJ...650.1070D}. The F160W ($\sim H$-band) absolute magnitude of the counterpart assuming 3.8\,kpc and A$_{\rm V}=7.6$ would be $\sim$5, similar to known counterparts \citep{2022MNRAS.513.3550C}. Placing it any further (up to the other extreme of 11\,kpc) and behind higher extinction \citep[up to the total Galactic extinction on that sight-line of A$_{\rm V}\sim60$,][]{2011ApJ...737..103S} pushes the source towards being intrinsically luminous and blue, inconsistent with known magnetar counterparts. It could feasibly then be a massive star, but given the co-location with the X-ray source, would presumably have to be a non-interacting companion to the magnetar, where the binary has been given a large natal kick without being unbound. The SNR has an estimated age of $\sim$12-14\,kyr (assuming d$=8-9$\,kpc), and this would only decrease if placed at a lower distance \citep{2017ApJ...847..121A,2021ApJ...923...15S}. We proceed under the assumption that the SNR is at the lower end of the distance estimates, and that 1RXS\,J1708 was born in this supernova, but acknowledge the significant uncertainties in this interpretation.

\section{Discussion}\label{sec:discuss}
\subsection{Magnetar natal kicks}
\begin{figure}
\centering
\includegraphics[width=0.49\textwidth]{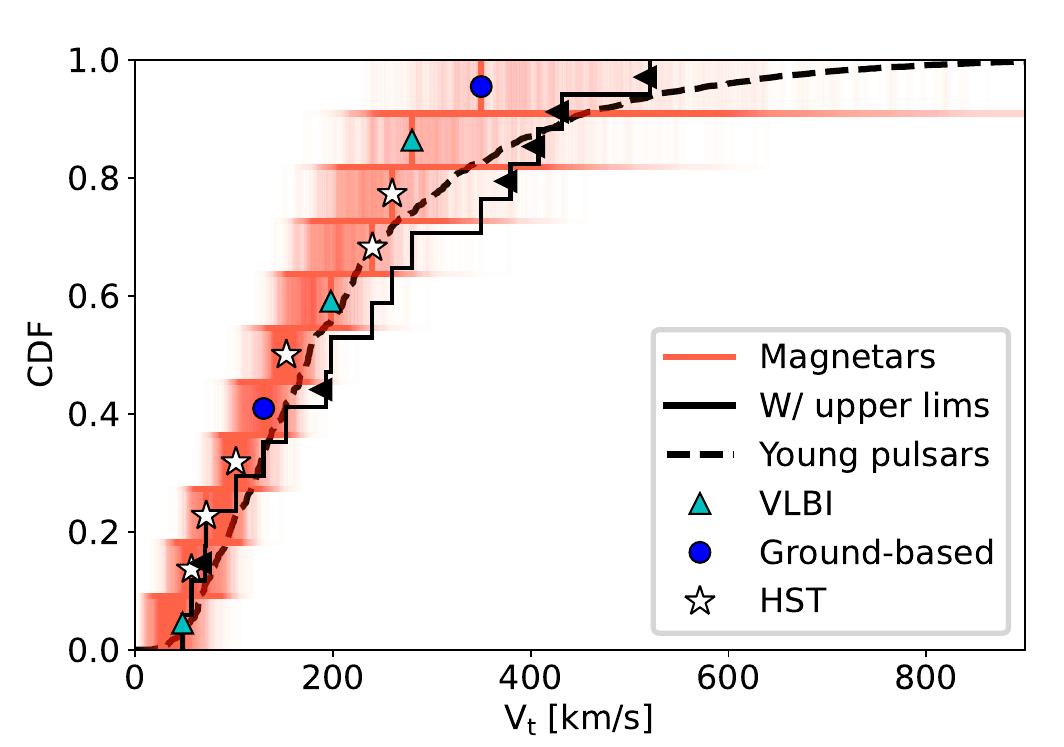}
\caption{The peculiar transverse velocity distribution (red) of all magnetars with a velocity measurement (i.e. transverse velocities after correction for an assumed distance and hence local standard of rest). Each step in the distribution is labelled according the method used to obtain the best available proper motion. One thousand random samples of the velocities, within their uncertainties (including distance uncertainty, see Table \ref{tab:mu}), are also shown. The solid black line represents the magnetar distribution with the six upper limits listed in Table \ref{tab:mu}; the steps corresponding these upper limits are marked with left-facing triangles. The transverse velocity distribution of pulsars with age $<$10\,Myr \citep[the fiducial distribution of][]{2025ApJ...989L...8D}is shown with a dashed black line.}
\label{fig:vt}
\end{figure}

\begin{figure*}
\centering
\includegraphics[width=0.49\textwidth]{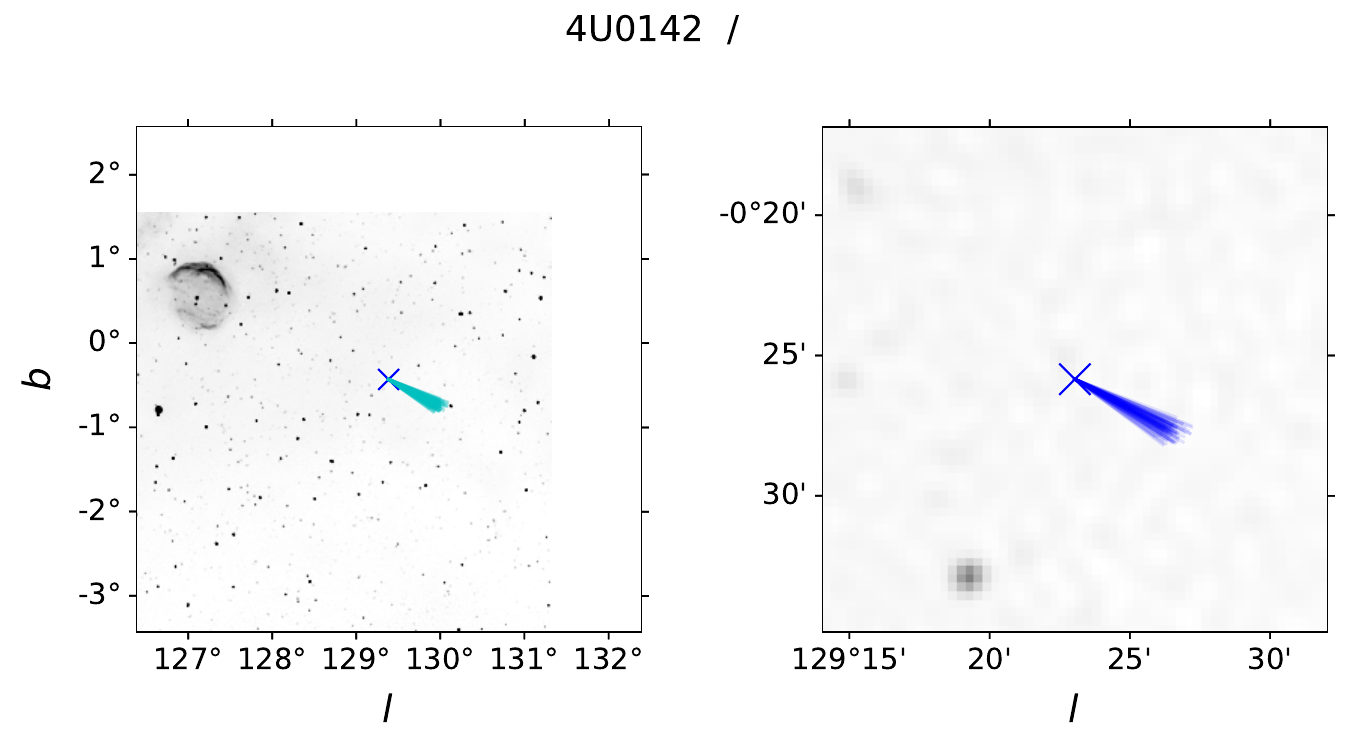}
\includegraphics[width=0.49\textwidth]{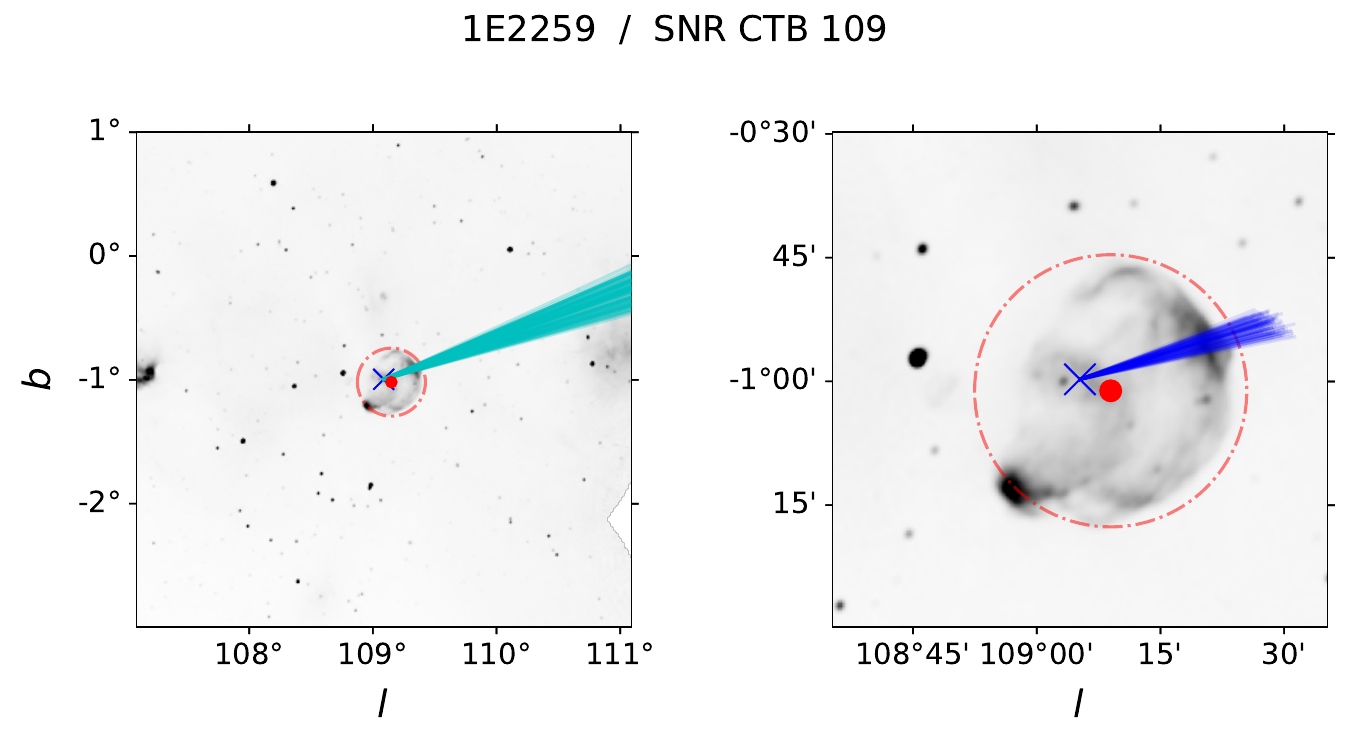}
\includegraphics[width=0.49\textwidth]{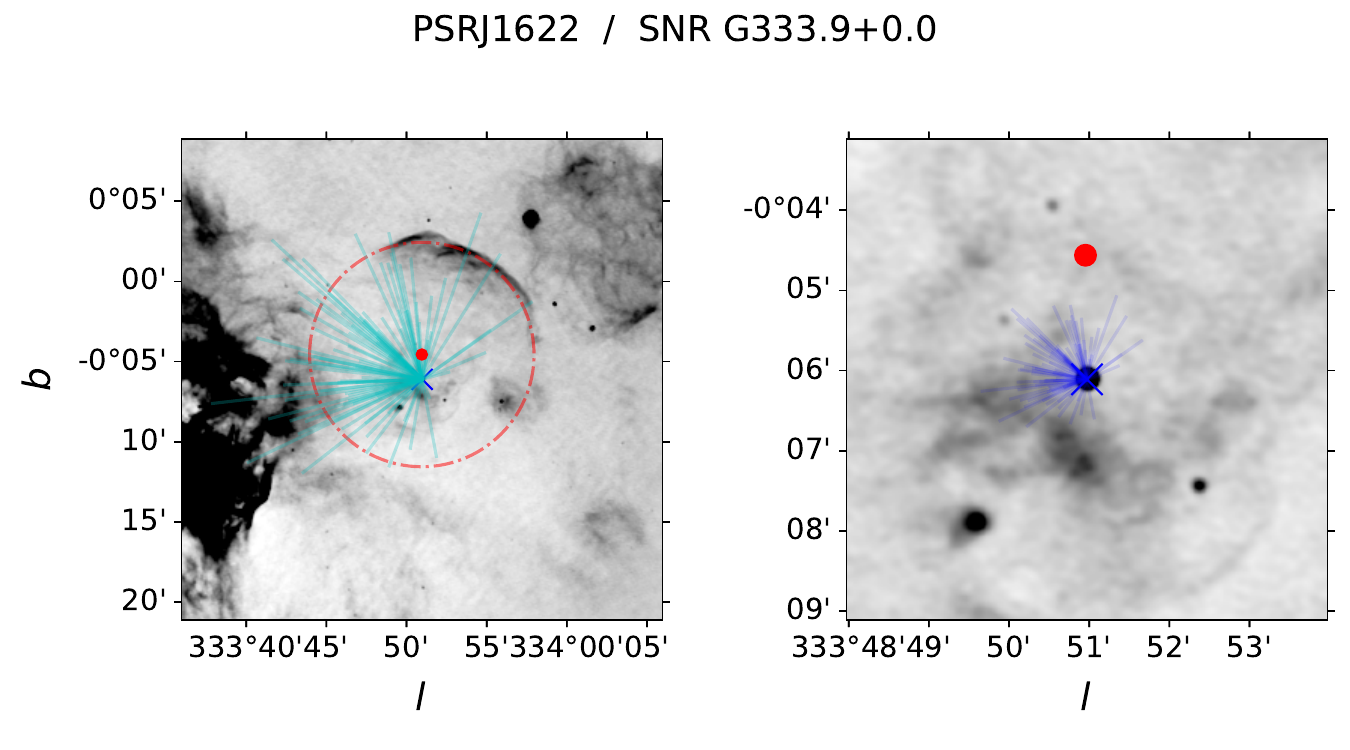}
\includegraphics[width=0.49\textwidth]{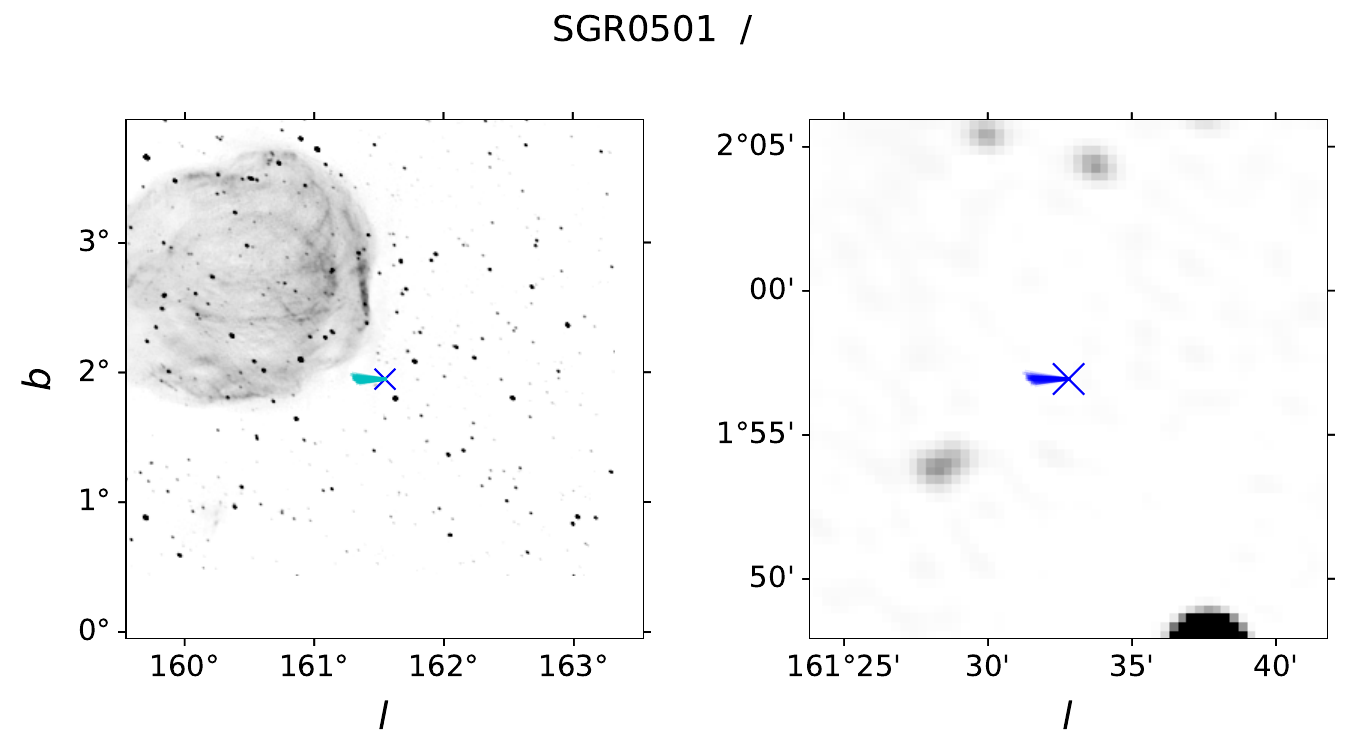}
\includegraphics[width=0.49\textwidth]{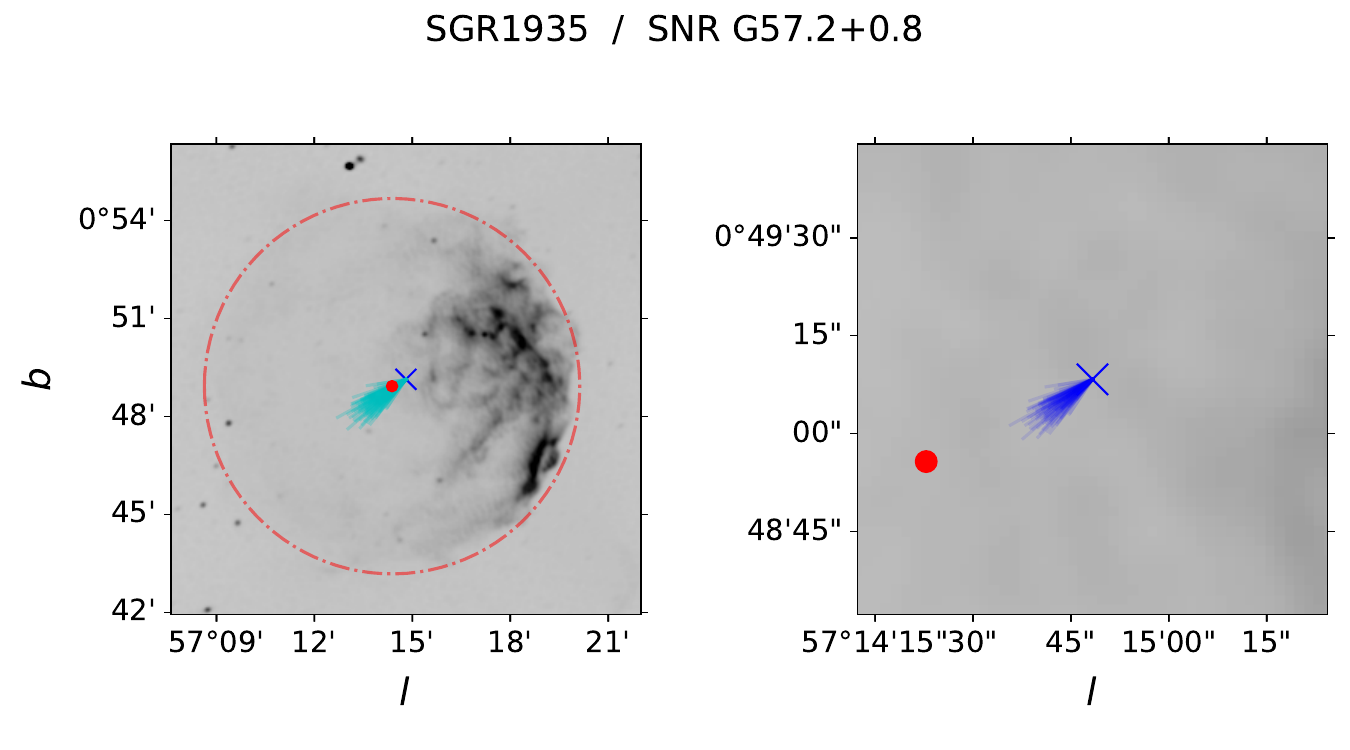}
\includegraphics[width=0.49\textwidth]{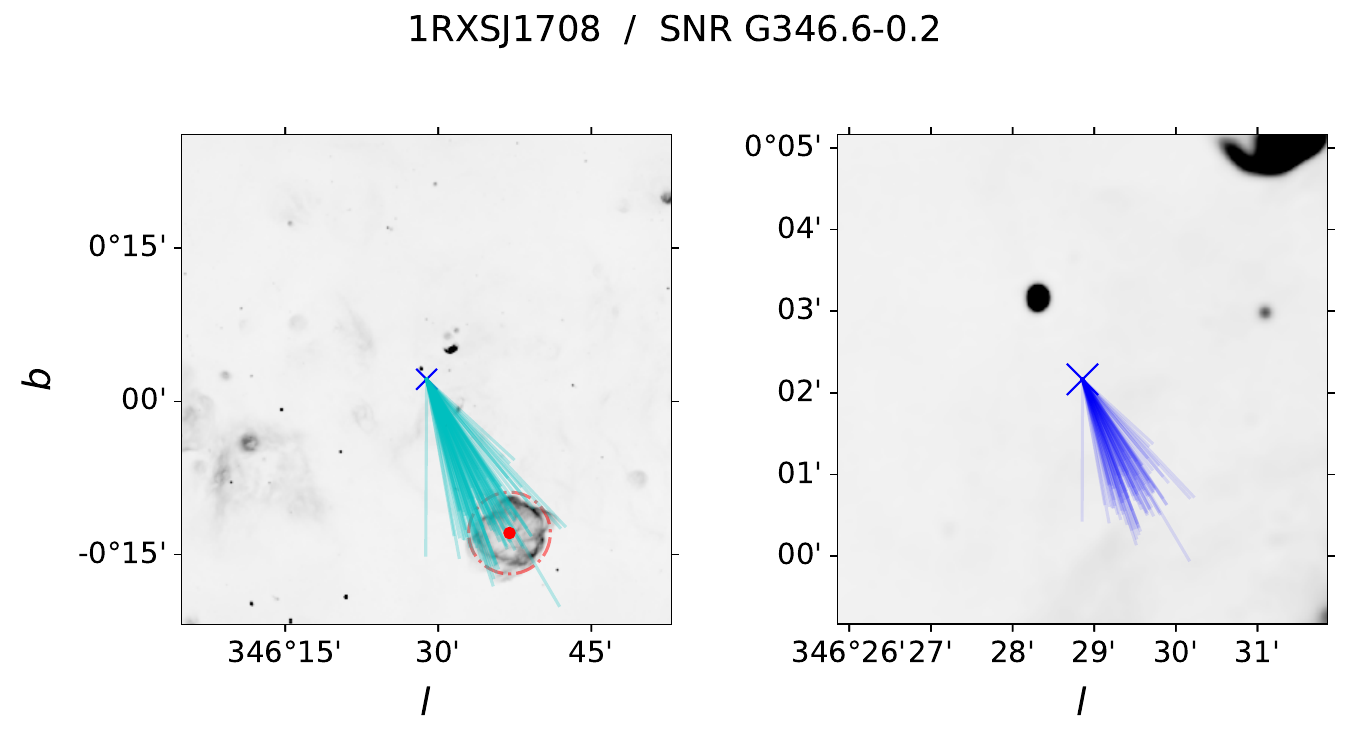}
\caption{The past trajectories of the magnetars for which we have measured a new/improved proper motion in this work. In each case we randomly sample 100 ($\mu_{\alpha}$cos($\delta$), $\mu_{\delta}$) pairs from the measured proper motions and their uncertainties, assuming they are Gaussian. The lengths of the past trajectory vectors correspond to (i) 10 times the characteristic age, in cyan, and (ii) characteristic ages, in blue. The radial extent of SNRs which are plausibly associated with a magnetar are indicated with dashed red lines, and the approximate centres of these SNRs are marked with a red dot. The background radio images are from the 1.4\,GHz Canadian Galactic Plane Survey \citep[4U0412, 1E2259 and SGR0501,][]{2003AJ....125.3145T} and the MeerKAT 1.3\,GHz Galactic Plane Survey \citep[PSRJ1622, SGR1935 and 1RXSJ\,1708,][]{2024MNRAS.531..649G}. No suitable SNR birth site can be identified for 4U0412 and SGR0501, analogous searches for stellar clusters in optical imaging have also produced null results for these objects \citep{2013ApJ...772...31T,2025AandA...696A.127C}.  }
\label{fig:traject}
\end{figure*}

The magnetar and young pulsar velocity distributions shown in Figure \ref{fig:vt} appear similar, with a possible dearth of magnetars at the high velocity end. We carry out Anderson-Darling tests between 1000 realisations of the magnetar distribution, sampling from their proper motion uncertainties, and the transverse velocity distribution of young pulsars \citep{2025ApJ...989L...8D}. We only consider magnetars with a measured proper motion, since the upper limits (corresponding to non-detections/non-identifications) may instead be cases where the counterpart was below the detection threshold of the image. To avoid negative v$_{\rm t}$ values we perform log-normal sampling of the velocities. P-values $>0.05$ are found in only 3\% of the realisations. This formally rejects the null hypothesis that the magnetar velocities are drawn from the same underlying distribution as pulsars. The 16th, 50th and 84th percentiles of the magnetar velocity distributions yield v$_{t}=154_{-86}^{+141}$\,km\,s$^{-1}$. We repeat the process, each time removing the fastest 1\% of the young pulsar distribution. We reach 5\% of realisations with $p>0.05$ once the fastest $\sim$15\% of pulsars have been removed. 
Similar to \citet{2024ApJ...971L..13D}, we therefore conclude that there is tentative evidence for a difference between the distributions, although this tension could plausibly be lessened by magnetar distance revisions. For example, we find that any distance revisions which gives just two magnetars in the sample a velocity of $>500$\,km\,s$^{-1}$ would be sufficient for the samples to become statistically indistinguishable. 

Given our selection criteria, slow-moving magnetars which are not particularly red or variable (or happen to have similar magnitudes in the relevant epochs) may have been missed. Our selection may therefore be biased towards faster moving objects. For example, the source in the error circle of SWIFTJ1834 appears to be moving against the bulk motion of objects in the field (Figure \ref{fig:cp_ident_no}). However, its displacement across the 5 year baseline is only $\sim 2\sigma$ outside the scatter of all sources in the field. Given that the source is also not anomalous in colour and variability, we cannot yet confidently claim it as the counterpart. Further imaging on longer temporal baselines could determine whether this source truly has a discrepant proper motion for its field. However, even in the cases where we only have an upper limit on the velocity (assuming the counterpart was detected at all), the magnetars would need to be close to these upper limits to extend the magnetar distribution into the pulsar high velocity tail, as shown in Figure \ref{fig:vt}.

Conversely, for high proper motions $>15$\,mas\,yr$^{-1}$, over baselines of $>5$ years, the object could have moved sufficiently that it is not recognised as the same object. A standard matching radius of 75\,mas is used to identify the common sources between images, but is extended by a factor of 2-3 inside the X-ray error circles to account for the possibility of high proper motion objects here. However, very high proper motion objects could have moved outside the X-ray error circle entirely, as the error circles are static and typically from observations taken several years prior to the first HST observations \citep{2014ApJS..212....6O}. This is a potential bias against the detection of very high velocity objects. To investigate, we visually inspect the images of every magnetar, blinking the images between epochs to check for high proper motions and/or objects leaving the X-ray error circle. This is effective because large changes in position are obvious by eye (e.g. the motion of 1RXSJ\,1708, with $\mu \sim$13\,mas\,yr$^{-1}$, is clearly visible when the images are blinked). We find no high-proper motion objects in the vicinity of the magnetar X-ray localisations, aside from those already identified as counterparts with the default selection criteria. We therefore conclude that a potential selection bias against very high v$_{\rm t}$ objects has not played a role in this sample. 

Finally, we investigate whether there is any correlation between magnetar magnetic fields and their transverse velocities, which might provide insight into their formation channels. Eleven magnetars have transverse velocity (Table \ref{tab:mu}) and magnetic field strength measurements \citep{2014ApJS..212....6O}. Notwithstanding the large velocity uncertainties, a Spearman rank correlation test yields a p-value of 0.36, indicating no evidence for a correlation between these two parameters.


\subsection{Kinematic versus characteristic ages}
For the sample of 11 magnetars with proper motion measurements, we collate information about their characteristic ($P$/$\dot{P}$ spin-down) age $\tau_{\rm c}$ in Table \ref{tab:age}. Spin-down ages assume a braking index $n=3$ \citep[e.g.][]{2015PhRvD..91f3007H}, that the magnetic field is dipolar and constant, and that $P_{\rm init} << P$. These assumptions may not hold, particularly for magnetars, where the excess luminosity above that expected from spin-down is attributed to the decay of the magnetic field \citep{1995MNRAS.275..255T}. Therefore, $\tau_{\rm c}$ may be an unreliable age indicator, and is an upper limit unless there are P-$\dot{P}$ variations due to variability during outbursts \citep{2003ApJ...596L..71K,2013ApJ...770L..23M}.

We also list in Table \ref{tab:age} the angular offset to any SNR or stellar cluster association, where available, and report a kinematic age $\tau_{k}$. The kinematic age uncertainties are propagated solely from the uncertainty on the proper motion, the uncertainty on the point of origin within the SNR or cluster is not quantified. Kinematic ages offers an independent constraint. They are defined as the time it would have taken, given the observed proper motion, to have travelled from the (estimated centroid of the) birth site to the current position. We further list the ages of associated SNRs in Table \ref{tab:age}, where available.

\begin{figure*}
\centering
\includegraphics[width=0.99\textwidth]{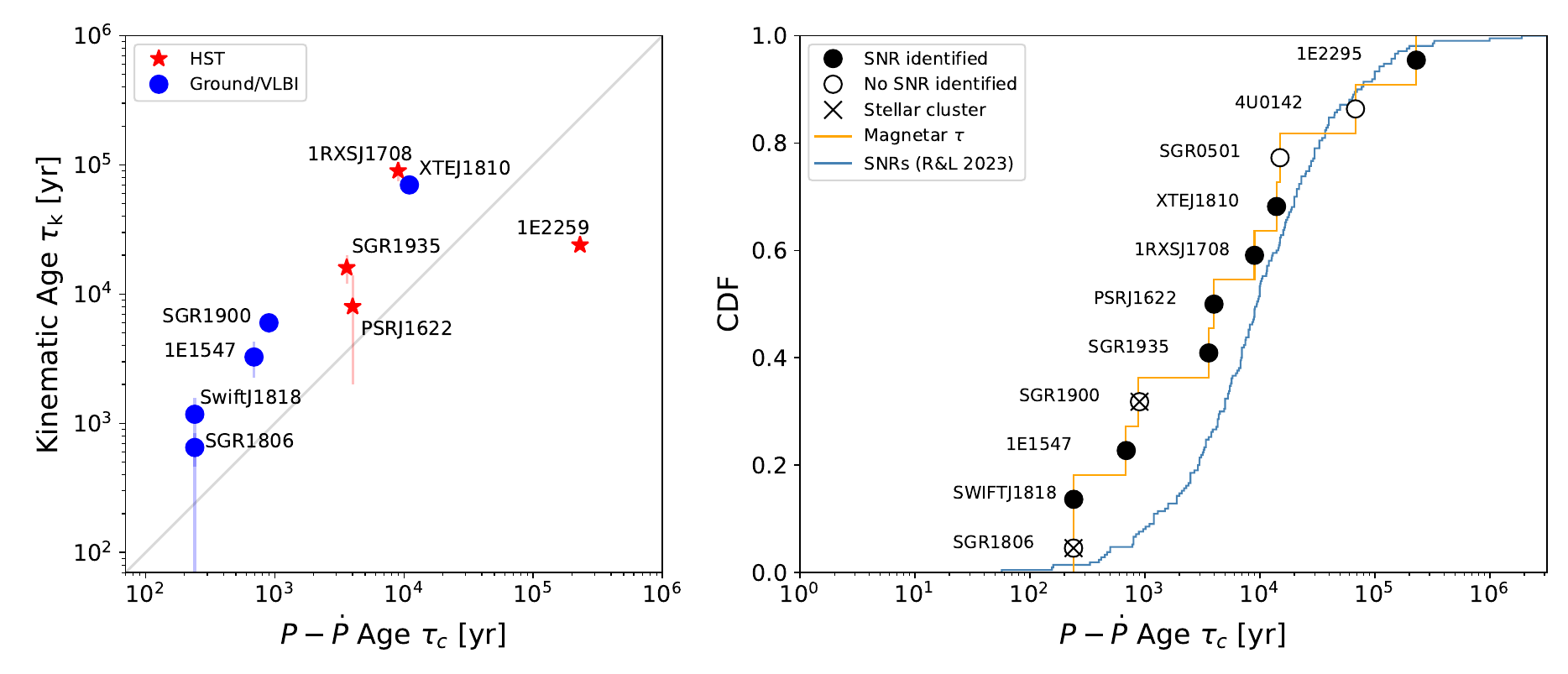}
\caption{Left panel: kinematic versus characteristic (spin-down) ages, for the subset of nine magnetars which have (i) a proper motion measurement and (ii) a SNR or stellar cluster along on the past trajectory (see Table \ref{tab:age}). The grey diagonal line represents $\tau_{c} = \tau_{k}$. The characteristic ages $\tau_{\rm c}$ are upper limits. The kinematic ages $\tau_{\rm k}$ assume an origin in the approximate centre of the associated SNR or cluster, and their uncertainties are propagated solely from the uncertainty on the proper motion. The SNR associations for XTEJ1810 and 1RXSJ1708 should be treated with particular caution due to the large discrepancy between the characteristic/SNR ages and the kinematic ages. Right panel: the cumulative distribution function (CDF) of characteristic ages for the 11 magnetars which have a proper motion measurement (not including the Galactic centre object SGR1745). Each object is labelled according to whether it has a SNR/stellar cluster association or no candidate birth site. Also shown is the cumulative distribution of Galactic SNR ages \citep{2023ApJS..265...53R}.}
\label{fig:kages}
\end{figure*}

In the left-hand panel of Figure \ref{fig:kages}, we plot $\tau_{\rm c}$ versus $\tau_{\rm k}$ for each case where a birth site has been identified. In every case the discrepancy is in the direction of the spin-down age underestimating the true age, with the sole exception of 1E2259. The mean shift (excluding 1E2259) is a factor of 4$\pm$1 increase over the characteristic age. In general, characteristic ages are not reliable because magnetar spin-down rates are known to vary, the magnetic field is not constant, and is known not to be a simple dipole \citep[e.g.][]{2009MNRAS.399L..44E}. However these effects are typically expected to act in the direction of decreasing the true age with respect to the characteristic age, so the result shown in Figure \ref{fig:kages} is surprising, with the exception of 1E2259 which stands out as the only example following the expected behaviour. 

Figure \ref{fig:kages} demonstrates the value of kinematic ages as an independent way to determine magnetar and SNR ages simultaneously, ultimately constraining the physics governing the spin-down evolution of magnetars \citep[e.g.][]{2012ApJ...761...76T,2026arXiv260206615L}. However, the kinematic ages are more uncertain than listed in Table \ref{tab:age} and shown in Figure \ref{fig:kages}, where the uncertainties are propagated solely from the proper motion uncertainties. In addition to this, there are the uncertainties of (i) where in the SNR or cluster the magnetar originated, which may not necessarily be the geometric centre as assumed and (ii) the reliability of the birth site association - we may be incorrectly assigning the birth site. In at least two cases (1RXSJ1708 and XTE J1810), there is a large discrepancy between the SNR age and kinematic age, which is challenging to explain and calls these associations into question. Furthermore, SGR1900 and SGR1806 have cluster associations, and there is a known bias against SNR detections in clusters (see the next subsection). Therefore, the true birth site may be closer than the cluster centroids assumed for the kinematic ages. If we assume that these four objects have spurious kinematic ages and remove them from Figure \ref{fig:kages}, the mean shift reduces to a factor of 3$\pm$1, and the significance of the result is further reduced when the uncertainty in the point of origin within the SNR is considered (e.g. SGR1935 need not have originated at the geometric centre of SNR G57.2+08). On the other hand, if the SNRs associated with magnetar formation have already faded, given the distribution of SNR ages (Figure \ref{fig:kages}), this might also suggest a true age which is larger than the characteristic age. To quantify these uncertainties, a detailed study of the observational biases at different wavelengths, the probability of chance alignments, and the evolution of SNRs throughout the Galactic plane would be needed.

If the result suggested by Figure \ref{fig:kages} were to hold, it would imply fewer young neutron stars in the Galaxy than suggested by their characteristic ages, which has implications for their birth rates and the fraction of neutron stars which are born as magnetars \citep[][]{2019MNRAS.487.1426B,2022MNRAS.509..634J,2025ApJ...986...88S,2026arXiv260116159P}. For example, in this sample, increasing the characteristic ages by a factor of three reduces the number of magnetars younger than 2\,kyr by a factor of 2 and those younger than 20\,kyr by a factor of 1.3. This would imply a corresponding reduction in the fraction of neutron stars born as magnetars, a reduction in the completeness of the Galactic magnetar sample, or some combination thereof. However, we stress again that these conclusions depend entirely on both the counterpart and birth site associations being robust.

\begin{table*}
	\centering
	\caption{Characteristic ages $\tau_{\rm c}$ \citep[from the McGill catalogue and references therein][]{2014ApJS..212....6O}, birth sites, and the approximate angular offsets to (the centre of) the proposed birth site. The six above the horizontal line have a HST derived proper motion; no birth site has been found for SGR0501 and 4U0142. Kinematic ages $\tau_{\rm k}$ are listed for the remaining 5 magnetars with a proper motion constraint and candidate birth site (see Section \ref{sec:vt} and Figure \ref{fig:vt}). All values are approximate, as we can cannot determine the precise place of origin in a cluster, or the exact centre of a SNR remnant (due to non-symmetric expansion). The final columns list supernova remnant ages $\tau_{\rm SN}$ (based on the size) where available, and their references.}%
	\label{tab:age}
	\begin{tabular}{lllp{3.5cm}llll} 
		\hline
		  \hline
            Magnetar	& $\tau_{\rm c}$  &	Birth site 	& Birth site ref(s). & Ang. sep. & $\tau_{\rm k}$ & $\tau_{\rm SN}$ & $\tau_{\rm SN}$ ref. \\
            	&  [kyr] &	 	&  & [arcmin] & [kyr] & [kyr] &  \\
            \hline						
            PSRJ1622	&	4	&	SNR G333.9$+$0.0	&	\citet{2012ApJ...751...53A}  	&	1.5	&	8$\pm$6 &	$<$6 & \citet{2010ApJ...721L..33L}	\\
            1E2259	&	230	&	SNR CTB 109	&	\citep{2013ApJ...772...31T}  	&	3	&	24$\pm$2 & 12.7$\pm$0.8 & \citet{2020ApJS..248...16L}		\\
            4U0142	&	68	&	?	&	\citep{2013ApJ...772...31T}  	&	-	&	- &	 - & - \\
            SGR0501	&	15	&	?	&	\citep{2025AandA...696A.127C}  	&	-	&	- &	 - & - 		\\
            SGR1935	&	3.6	&	 SNR\,G57.2+0.8	&	\citet{2022ApJ...926..121L}  	&	0.6	&	16$\pm$4 & $\gtrsim$16	& \citet{2020ApJ...905...99Z}	\\
            1RXSJ\,1708 $\dagger$ 	&	9	&	SNR\,G346.6-0.2	&	This work  	&	20	&	 90$\pm$15 & 14$\pm$2 & \citet{2021ApJ...923...15S}		\\
            \hline											
            1E1547	&	0.69	&	SNR G327.24–0.13	&	\citet{2007ApJ...666L..93C}  	&	0.5	&	3$\pm$1 &	- & -	\\
            SGR1806	&	0.24	&	Stellar cluster	&	\citet{2012ApJ...761...76T}  	&	0.07	&	0.65$\pm$0.19 &	- & -		\\
            SGR1900	&	0.9	&	Stellar cluster	&	\citet{2012ApJ...761...76T}  	&	0.33	&	6$\pm$1 &	- & -		\\
            XTE\,J1810 $\dagger$ 	&	11	&	SNR G11.0-0.0	&	\citep{2007ApJ...662.1198H}  	&	19	&	70$^{+8}_{-10}$ &	$\lesssim$ 3  & \citet{2020MNRAS.498.3736D}		\\
            SWIFTJ1818	&	0.24	&	SNR	&	\citet{2023ApJ...943...20I}  	&	0.2$\star$	&	1.20$_{-1.15}^{+0.40}$ &	- & -	\\
            \hline	
	\end{tabular}\newline
    $\star$ - Judged by eye from Fig. 7 of \citet{2023ApJ...943...20I}. $\dagger$ - These magnetar-SNR associations should be treated with caution due to the large discrepancy between the characteristic/kinematic ages, and the SNR age.
    \end{table*}

\subsection{Magnetars without SNR or cluster associations}
A parent SNR or stellar cluster has not been identified in every case. We ask whether this is surprising, given the age of the magnetar, and assuming that the characteristic age is representative (within some scatter as described above). In the right hand panel of Figure \ref{fig:kages}, we plot the cumulative distribution of $\tau_{\rm c}$ for the 11 magnetars with a proper motion and hence an opportunity to determine the birth site. Each step in the distribution is labelled by the magnetar name and whether a SNR association has been made. Also shown is the age distribution of Galactic SNRs \citep{2023ApJS..265...53R}.  All 11 magnetars lie in the same age range as Galactic SNRs, and 7 have at least a tentatively associated SNR which is consistent with the past trajectory. Although not included here due to the likely influence of Sgr A$^{\star}$ on its motion, SGR1745 has also been linked with a SNR \citep{2015MNRAS.453..172P,2017ApJ...838...12Y}. 

A further two (SGR1900 and SGR1806) are not associated with a SNR, these both have ages at the lower end of the Galactic SNR age distribution. They are, however, linked with massive stellar clusters \citep{2012ApJ...761...76T}. This could reflect different formation channels: for example, one channel could arise from progenitors which are ejected and travel far outside their birth clusters before going supernova \citep[e.g.][]{2011MNRAS.414.3501E}, while in the other the progenitor remains in situ but produces weak SNe. However, there are known biases against SNR detection in clusters due to to the merging of individual remnants in dense environments \citep[e.g.][]{2017MNRAS.468.2757V}, confusion with the diffuse background/HII region emission \citep{2017A&A...605A..58A,2025A&A...693A.247A,2019JApA...40...36G} and the small angular sizes of young SNRs in dense environments \citep{2021MNRAS.504.1536V}.

Finally, SGR0501 and 4U0142 are the only two magnetars with a well constrained past trajectory but no obvious SNR or parent stellar cluster. They are among the oldest objects in the sample, so plausibly an associaed SNR may have already faded, although SNRs with similar ages are known. If the SNR HB9 were the origin of SGR0501 (neglecting for a moment the direction of the proper motion vector), the kinematic age would be $\sim$10$^{6}$ years, in the extreme tail of the SNR age distribution. Given the spin-down age of $\sim$15\,kyr, this would be by far the largest discrepancy in the left panel of Figure \ref{fig:kages}. Although we show characteristic ages in the right panel of Figure \ref{fig:kages}, the true (kinematic) age of 1E2259 is a factor of $\sim$10 younger, so the detection of the associated SNR is consistent known SNR ages, in contrast with the 4U0142 and SGR0501 non-detections. If the association of 1RXSJ\,1708 with SNR G346.6-0.2 holds, it must be a factor of $\sim$10 older than the characteristic age, which would also make SNR G346.6-0.2 one of the oldest SNRs known. 
 
Crucially, in the other cases where no SNR was identified (SGR1900 and SGR1806), massive star clusters were nevertheless present. Given that the lifetimes of even massive stars are at least 2 orders of magnitude longer than the characteristic ages of SGR0501 and 4U0142, and their positions in the Galactic anti-centre direction with relatively low extinction on their sight-lines, the lack of massive stars/clusters along their past trajectories is also surprising. Non-core-collapse channels such as accretion induced collapse \citep{1991ApJ...367L..19N} and low-mass neutron star mergers \citep{2013ApJ...771L..26G,2017ApJ...844L..19P} are plausible alternative progenitor channels \citep{2025AandA...696A.127C}. However, these interpretations are challenged by the low Galactic latitudes of SGR0501 and 4U0412 which should not necessarily be favoured by pathways involving white dwarfs (which are not confined to the disc) or neutron star mergers \citep[which can be kicked out of the plane,][]{}. Similarly, the explanation that their progenitor stars were ejected from their parent clusters, either by the supernova of a companion \citep[e.g.][]{1961BAN....15..265B,2019A&A...624A..66R} or dynamical interactions \citep[e.g.][]{1967BOTT....4...86P,2005A&A...437..247D} faces a similar problem, since massive runaway stars could travel hundreds or even thousands of parsecs (easily enough to escape the Galactic plane) in such a scenario before going supernova \citep[e.g.][]{2011MNRAS.414.3501E}.

Finally, we can ask whether the cluster-associated, SNR-associated and non-associated magnetars have distinct properties. In terms of velocities, the cluster-associated magnetars occupy the middle of the transverse velocity distribution, while 4U0142 and SGR0501 lie at the lower end. Although the sample size is small, we note that the two cluster-associated magnetars (SGR1806 and SGR1900) also have the strongest magnetic fields in the sample and are among the youngest, while 4U0142 and SGR0501 are among the oldest. This may be consistent with a scenario in which young SNRs in clusters are hard to detect, and older remnants have faded, as discussed above.

\section{Conclusions}\label{sec:conc}
In this paper, we searched for NIR counterparts to Galactic magnetars in repeat HST (and in one case single-epoch JWST) imaging. We re-discovered the counterparts of four magnetars (4U0142, 1E2259, SGR0501, SGR1935), reducing the uncertainties on their proper motions, and report new candidates for a further three (PSRJ1622, 1RXSJ\,1708, CXOUJ1647), measuring the first proper motions for PSRJ\,1622 and 1RXSJ\,1708. We find that magnetar characteristic ages may be underestimates of their true ages, based on SNR and stellar cluster associations. However, this results hinges on the reliability of these birth site associations.

We show that magnetar transverse velocities and birth sites are broadly similar to that of young pulsars, with the notable exceptions of SGR0501 and 4U0142 for which no candidate birth sites have yet been identified. This is consistent with the idea that magnetars evolve into other classes of neutron star such as X-ray dim isolated neutron stars \citep[e.g.][]{2023hxga.book..146B,2024MNRAS.534.1481G}, and searches for companion stars \citep{2022MNRAS.513.3550C}, which broadly match expectations for neutron stars in general. However, there is some evidence \citep{2024MNRAS.531.2379S} that magnetars may be preferentially formed in the core-collapse of merger products based on a lack of unbound companions \citep[see also][]{2012ApJ...761...76T,2019Natur.574..211S,2020MNRAS.495.2796S,2023Sci...381..761S}. We also show that tentative evidence is emerging for a dearth of high-velocity magnetars, compared with young pulsars. If real, this effect may arise from an as yet unidentified physical preference in the progenitor population which produces magnetars, and/or some difference in their post-formation velocity evolution. 

The faintness of the new counterparts we have identified in this work ($>$23.5 AB at 1.6$\mu$m), coupled with crowded sight-lines through the Galactic plane, demonstrates the strength of sensitive, high spatial resolution NIR observations for multi-wavelength neutron star studies. We have also shown that, while effective, magnetar counterpart selection based on colour, variability and proper motion may miss some objects and could be biased towards higher transverse velocities. For example, the discovery of the counterpart of CXOUJ1647 required a JWST spectral energy distribution. Future multi-filter and multi-epoch JWST studies will enable redder, fainter and higher spatial resolution observations. The constraints achievable depend on the signal-to-noise (S/N) ratio of the source, but for targets detected with S/N$=20$, we estimate that with observing baselines of 3 years, NIRcam observations can probe proper motions as small as 1\,mas\,yr$^{-1}$. At this level, parallaxes may even be measurable for nearby magnetars if the observations are coordinated carefully \citep[previously only possible for neutron stars within $\sim$100\,pc,][]{2002ApJ...576L.145W}. There are a range of potential applications, not only to magnetars, but any faint objects in the Galactic plane and Galactic centre for which kinematics are scientifically interesting \citep[e.g.][]{2023ApJ...950..101L}.

\begin{acknowledgements} 
We thank the referee for their report which improved the clarity and completeness of this paper. The authors also thank Nanda Rea for insightful discussions. \\
AAC and AB acknowledge support through the European Space Agency (ESA) Research Fellowship in Space Science.
JDL acknowledges support from a UK Research and Innovation Future Leaders Fellowship (grant references MR/T020784/1 and UKRI1062).
NRT is supported by STFC Consolidated grant ST/W000857/1.\\

This research is based on observations made with the NASA/ESA Hubble Space Telescope obtained from the Space Telescope Science Institute, which is operated by the Association of Universities for Research in Astronomy, Inc., under NASA contract NAS 5–26555. These observations are associated with program 17927.

This work is based in part on observations made with the NASA/ESA/CSA James Webb Space Telescope. The data were obtained from the Mikulski Archive for Space Telescopes at the Space Telescope Science Institute, which is operated by the Association of Universities for Research in Astronomy, Inc., under NASA contract NAS 5-03127 for JWST. These observations are associated with program GO 1905.

This work made use of data from the European Space Agency (ESA) mission {\em Gaia} (\url{https://www.cosmos.esa.int/{\em Gaia}}), processed
by the {\em Gaia} Data Processing and Analysis Consortium (DPAC, \url{https://www.cosmos.esa.int/web/{\em Gaia}/dpac/consortium}). Funding for the DPAC has been provided by national institutions, in particular the institutions participating in the {\em Gaia} Multilateral Agreement.\\

We acknowledge use of the McGill magnetar \citep{2014ApJS..212....6O}, Green \citep{2025JApA...46...14G}, and University of Manitoba SNRcat \citep{2012AdSpR..49.1313F}\footnote{\url{http://snrcat.physics.umanitoba.ca}} supernova remnant catalogues.

This research has made use of the SVO Filter Profile Service "Carlos Rodrigo", funded by MCIN/AEI/10.13039/501100011033/ through grant PID2023-146210NB-I00 \citep{2012ivoa.rept.1015R,2020sea..confE.182R,2024A&A...689A..93R}.

This work made use of {\sc ipython} \citep{2007CSE.....9c..21P}, {\sc numpy} \citep{2020Natur.585..357H}, {\sc scipy} \citep{2020NatMe..17..261V}; {\sc matplotlib} \citep{2007CSE.....9...90H}, and {\sc astropy} (\url{https://www.astropy.org}), a community-developed core Python package for Astronomy \citep{astropy:2013, astropy:2018}. We have also made use of the python modules {\sc dustmaps} \citep{2018JOSS....3..695G} and {\sc extinction} \citep{barbary_kyle_2016_804967}.

\end{acknowledgements} 

%
\bibliographystyle{aa} 
\bibliography{mag_mu} 
%

\end{document}